\documentclass[%
twocolumn,reprint,superscriptaddress,
showpacs,preprintnumbers,
bibnotes,
amsmath,amssymb,
aps,prb
]{revtex4-2}

\usepackage{ulem} % allow \normalem
\usepackage{siunitx}
\usepackage{graphicx}% Include figure files
\usepackage{dcolumn}% Align table columns on the decimal point
\usepackage{bm}% bold math
\usepackage{color}
\usepackage{subeqnarray}
\usepackage{cases}
\usepackage{booktabs}
\usepackage{upgreek}
\usepackage{geometry}
\usepackage{type1cm}
\usepackage[marginal]{footmisc}
\usepackage{appendix}
\usepackage{hyperref} % link
\usepackage{bm} % link
\usepackage{hyperref}% add hypertext capabilities, doi or url

\begin{document}
	
	\title{Exploration of magnetoelastic deformations in spin-chain compound CuBr$_2$}
	
	\author{Biaoyan~Hu}
	\email{hubiaoyan@quantumsc.cn}
	\affiliation{International Center for Quantum Materials, Peking University, Beijing, China}
	\affiliation{Quantum Science Center of Guangdong-Hong~Kong-Macao Greater Bay Area, Shenzhen, China}
	\author{Yingying~Peng}
	\affiliation{International Center for Quantum Materials, Peking University, Beijing, China}
	\author{Xiaoqiang~Liu}
	\affiliation{International Center for Quantum Materials, Peking University, Beijing, China}
	\author{Qizhi~Li}
	\affiliation{International Center for Quantum Materials, Peking University, Beijing, China}
	\author{Qiangqiang~Gu}
	\affiliation{International Center for Quantum Materials, Peking University, Beijing, China}
	\author{Matthew~J.~Krogstad}
	\affiliation{Materials Science Division, Argonne National Laboratory, Lemont, IL, USA} % 一说是 Argonne, IL
	\author{Raymond~Osborn}
	\affiliation{Materials Science Division, Argonne National Laboratory, Lemont, IL, USA} % 一说是 Argonne, IL
	\author{Takashi~Honda}
	\affiliation{Institute of Materials Structure Science, High Energy Accelerator Research Organization (KEK), Tokai, Ibaraki, Japan}
	\author{Ji~Feng}
	\affiliation{International Center for Quantum Materials, Peking University, Beijing, China}
	\author{Yuan~Li}
	\affiliation{International Center for Quantum Materials, Peking University, Beijing, China}
	
	\begin{abstract}

We investigate a spin-$\frac{1}{2}$ antiferromagnet, CuBr$_2$, which has quasi-one-dimensional structural motifs. The system has previously been observed to exhibit unusual Raman modes possibly due to a locally deformed crystal structure driven by the low-dimensional magnetism. Using hard X-ray scattering and neutron total scattering, here we aim to verify a specific form of tetramerized lattice deformation proposed in the previous study. Apart from diffuse scattering signals which we can reproduce by performing a thorough modeling of the lattice's thermal vibrations, we do not observe evidence for a tetramerized lattice structure within our detection sensitivity. As a result, it is more likely that the unusual Raman modes in CuBr$_2$ arise from classical magnon-phonon hybridization, rather than from quantum spin-singlet-driven lattice deformation. 

	\end{abstract}
	\maketitle
	
	\section{Introduction}

Low-dimensional magnets command attention in scientific studies \cite{hess2007heat, lemmens2003magnetic, sheka2015curvature, vasiliev2018milestones, bethe1931theorie}. In these systems, long-range magnetic order weakens, giving rise to localized spin entanglement, including spin singlets, especially in spin-$\frac{1}{2}$ chain systems \cite{aplesnin2000static, kokalj2015antiferromagnetic, shu2018dynamical, sahling2015experimental}. Introducing frustration transforms the antiferromagnetic chain, resulting in spiral spin order and the formation of intriguing spin singlets \cite{haldane1982spontaneous, wellein1998peierls, becca2003tetramerization}, which interact with the Néel state, as supported by theoretical studies and experimental evidence \cite{nishimori1987competition, yamase2004competition}. 

The study of spin singlets is important, as they play a key role in forming the resonating valence bond (RVB) state, which is among the prominent explanations of high-temperature superconductivity \cite{anderson1987resonating, anderson1987resonatingPRL, anderson2004physics}. Looking into spin singlets may help us better understand correlated electronic materials and quantum magnetism. Spin singlets have been identified in various low-dimensional magnets \cite{ninomiya2003observation, deng2013coexistence, pasco2019tunable, danilovich2017spin}. Recent Raman studies on a low-dimensional magnet CuBr$_2$ have also suggested potential signs of spin singlets \cite{wang2017observation}. This study aims to verify the spin singlets in CuBr$_2$. 

CuBr$_2$ is a quasi-one-dimensional spin-$\frac{1}{2}$ frustrated antiferromagnet. It crystallizes in the monoclinic space group C$2$/m with lattice constants of $a\approx 7.2$~\AA, $b\approx 3.5$~\AA, $c\approx 7.0$~\AA\ and $\beta\approx 119.6^\circ$. CuBr$_2$ exhibits a N\'eel temperature ($T_{\rm N}$) of $73.5$ K \cite{wang2017observation, zhao2012cubr2, wang2018nmr, apostolov2022magnetic}. Below $T_{\rm N}$, CuBr$_2$ adopts a N\'eel state characterized by a spiral magnetic order, with a propagating wave vector of ${\bf q}_{\rm s}=(1, 0.235, 0.5)$ in reciprocal lattice units (r.l.u.) \cite{wang2017observation, lee2012investigation, lebernegg2013magnetism}. The spin rotation between adjacent Cu ions along chains is approximately 85$^\circ$ due to the competition between the nearest-neighbor ($J_1$) and the next-nearest-neighbor ($J_2$) interaction \cite{lee2012investigation, wang2017observation}.

\begin{figure}
		\includegraphics[width=3.2in]{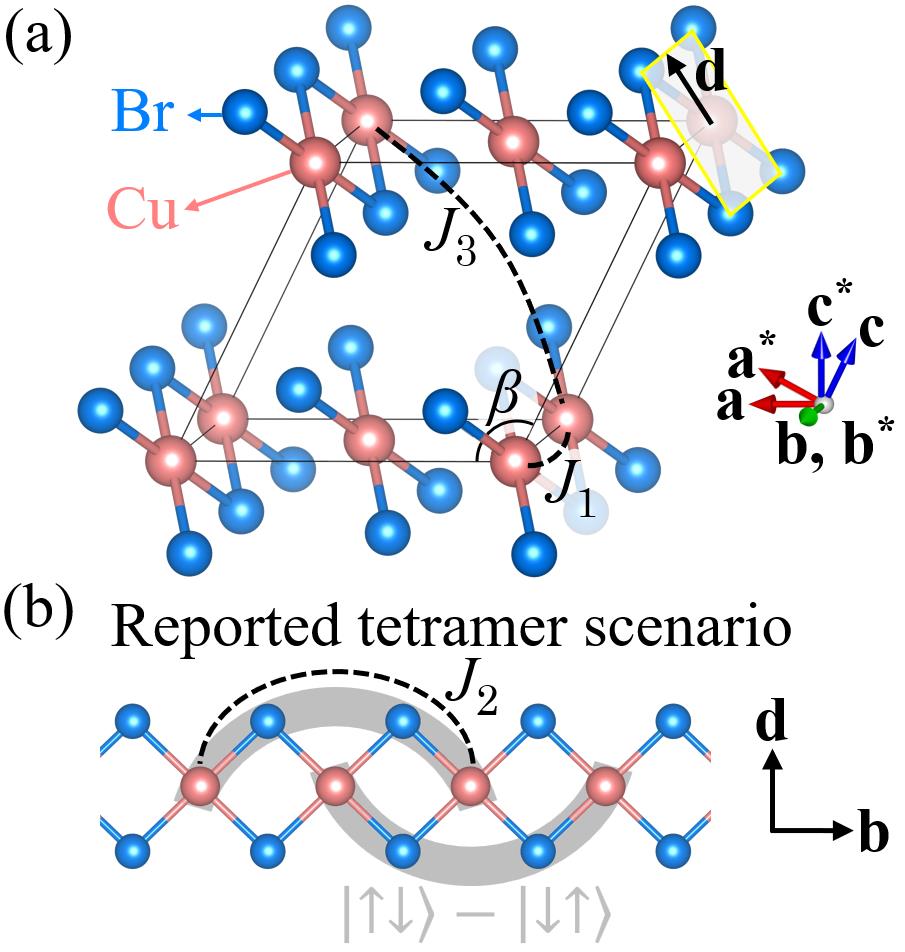}
		\caption{\label{CuBr2}
			(a) Crystal structure of CuBr$_2$. Here vector $\bf d$ is defined as shown, which is approximately parallel to (${\bf a}+{\bf c}$). 
			(b) The diagram of the reported tetramer scenario in the previous study \cite{wang2017observation}. The grey curve connected two Cu atoms is a spin singlet. In this scenario, two singlets may form a tetramer. }
	\end{figure}

The Raman scattering spectrum of CuBr$_2$ revealed signals intriguing non-monotonic variations with temperature at specific energies, exhibiting a pronounced maximum near the N\'eel temperature $T_{\rm N}$ \cite{wang2017observation}. Notably, these unusual modes did not align with the phonon energies at $\Gamma$ points, but coincided instead with the phonon energy at ${\bf Q}\approx0.25\ {\bf b}^*$ \cite{wang2017observation}. In an attempt to elucidate this phenomenon, two compelling explanations were proposed, one of which focusing on the hypothetical singlets. Given that antiferromagnetic $J_2$ is the dominant spin interaction among the three primary interactions ($J_1$, $J_2$, and $J_3$) \cite{lee2012investigation}, it has been proposed in the previous study that spin singlets might emerge within the next-nearest neighboring Cu pairs along chains \cite{wang2017observation}, as represented in Fig.~\ref{CuBr2} (b). Consequently, spin interactions might revise on the lattice, leading to the tetramerized lattice deformation, which provided an explanation for the observed magnetoelastic coupling effect in the Raman experiments \cite{wang2017observation}. 

To experimentally verify the existence of tetramers in CuBr$_2$, scattering experiments are undertaken. These experiments involve an analysis of both the momentum distribution of the intensity and the pair distribution function (PDF). By analyzing momentum distribution intensity and studying the PDF, we aim to provide evidence and insights into the potential tetramerized deformation in CuBr$_2$. These experiments would enhance our grasp of quantum magnetism. 

We conduct a detailed hard X-ray scattering experiment on a pristine CuBr$_2$ single crystal to investigate the atomic displacements potentially induced by tetramers. Leveraging the force constants obtained from density functional theory (DFT), we calculate the thermal diffuse scattering (TDS). The TDS intensity is calculated and show agreement with the experimental data, offering no indication of the presence of tetramers. 

In order to explore the possibility of highly randomized tetramers, we perform an extensive analysis of the PDFs derived from neutron total scattering measurements on CuBr$_2$ powder. However, despite a thorough examination, we do not detect any distinctive variations in the nearest Cu-Cu pairs, thus reinforcing the conclusion that tetramers are not present in our study. 

These findings imply a higher likelihood that the unusual Raman modes observed in CuBr$_2$ result from classical magnon-phonon hybridization, rather than being driven by quantum spin-singlet-induced lattice deformation. 

\section{Methods}

\subsection{Sample preparation}

The growth of the CuBr$_2$ single crystal involve a controlled process of slow evaporation. Initially, a saturated solution of CuBr$_2$ is prepared and placed in a dry and stable environment at room temperature. With great care, the solution is allowed to gradually evaporate until it reaches complete dryness, resulting in the formation of high-quality sub-centimeter single crystals. Throughout this process, it is essential to strictly avoid any form of heating, as it could lead to the undesirable production of CuBr and compromise the purity of the sample.

In order to ensure the utmost accuracy of the PDF data, a procedure is employed to obtain high-quality CuBr$_2$ powder. The saturated CuBr$_2$ solution is subjected to a freeze-drying technique, which effectively precipitate the CuBr$_2$ compound in a crisp and sponge-like form. By gently crushing the sponge-like CuBr$_2$, powder is obtained with exceptional purity, well-defined crystallinity, and a satisfactory random orientation. These attributes are essential for the intended analysis and characterization of the material. 

\subsection{Hard X-ray scattering}

A hard X-ray scattering experiment is conducted on a CuBr$_2$ single crystal to unravel the fine atomic structure. X-rays are chosen as the primary investigative tool due to their precision in targeting and analyzing non-magnetic signals. 

In the hypothesis of minor lattice deformations, their influence on small momenta becomes negligible as long as these deformations are much smaller than the wavelength scale. Hard X-rays are essential for detecting subtle deformations and exploring large momentum transfers in this experiment, helping understand intricate structural changes in the crystal lattice. 

\begin{figure}
	\includegraphics[width=3.2in]{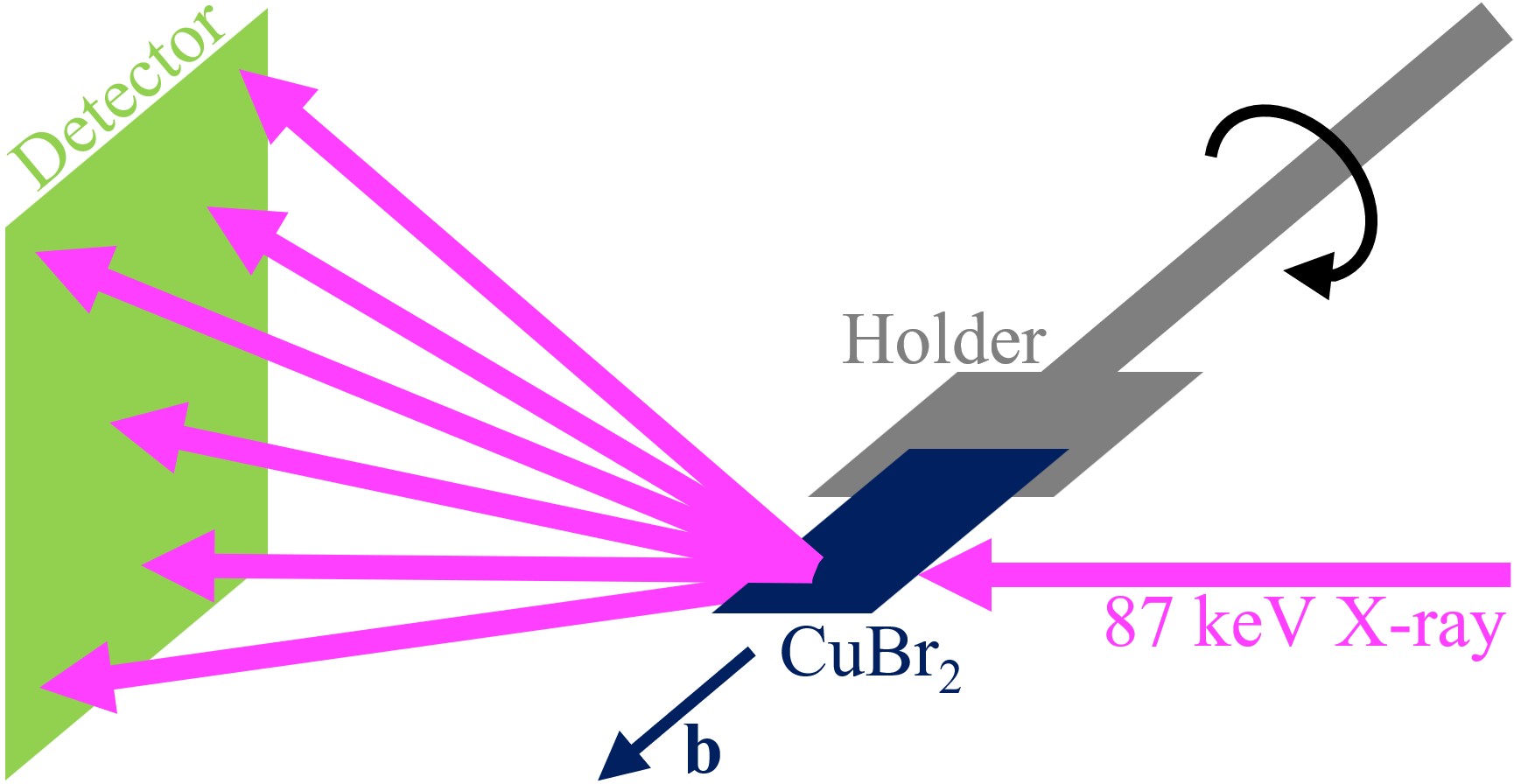}
	\caption{\label{HXS}
		Diagram of the hard X-ray scattering experiment. X-rays are transmitted through the CuBr$_2$ single crystal and scattered onto the detector screen. 
		The sample holder rotates 360° around $\bf b$ with the step of 0.1°. 
	}
\end{figure}

The experimental setup, illustrated in Fig.~\ref{HXS}, involves the penetration of 87-keV hard X-rays through the sample and subsequently scattered onto a rectangular detector screen. The detector array captures a specific section of the Ewald sphere corresponding to the sample's orientation. By rotating the sample, the Ewald sphere sweeps through the reciprocal space, generating a three-dimensional volume data set. Notably, the rotation axis is perpendicular to the initial X-ray direction. 

To adequately cover the regions surrounding the $\mathbf{b^*}$ axis in the reciprocal space, a deliberate slight angle is introduced between the crystal's $\mathbf{b}$ axis and the rotation axis. The data collection process involves performing measurements at 31 different temperatures, spanning from 30 to 300 K. At each temperature, the sample holder is rotated by 0.1° increments, completing a full 360° rotation.

Considering the hypothetical scenario depicted in Fig.~\ref{CuBr2}(b), the scattering intensity may exhibit superstructural signals which might manifest as a series of walls in the reciprocal space. For ${\bf Q} = h{\bf a^*} + k{\bf b^*} + l{\bf c^*}$, the potential superstructural signals might appear at $k=\cal{N}/\rm{4}, \cal{N}\in \mathbb{Z}$. To identify the potential superstructural signals, it becomes necessary to calculate the TDS. 

\subsection{Calculation of thermal diffuse scattering}

TDS is the coherent component of diffuse scattering, serving as a key element within it. TDS originates from the vibrations of atoms \cite{Warren1990, James1954, AlsNielsen2011elements} and can be expanded as \cite{Warren1990, baron2015introduction}: 
\begin{equation}\label{Stds=sumSp}
	S_{\rm TDS} = \sum_{p=1}^{\infty}S_p = \sum_{p=1}^{n}S_p + R_n,
\end{equation}
in which $S_p$ represents the $p$th-order TDS \cite{Warren1990}, corresponding to the scattering that involve the creation or annihilation of $p$ phonons \cite{baron2015introduction}. The remainder term $R_n$ is defined as $R_n\equiv\sum_{p=n+1}^{\infty}S_p$, encompassing the contributions from higher-order terms beyond $n$. 

To estimate the remainder term, we propose a model that incorporates random repositioning of each atom. This model provides a framework to calculate the remainder term $R_n$, which can be expressed as:
\begin{equation}\label{Rn}
	R_n\approx\frac{1}{\nu}\sum_{d=1}^{\nu}f_d^2\left(1-e^{-2W_d}\sum_{p=0}^{n}\frac{(2W_d)^p}{p!}\right). 
\end{equation}
Here, $d$ represents the serial number of the atom within a primitive cell, while $\nu$ denotes the total number of atoms in a primitive cell, which is 3 for CuBr$_2$. $W_d$ represents the Debye-Waller factor. $f_d$ corresponds to the atomic form factor, which employed in this work are sourced from the ionic data provided by Brown \textit{et al.} \cite{brown2006Intensity}. 

To determine the accurate values of $S_p$ and $W_d$, we utilize the force constant matrices obtained from DFT using Vienna \textit{ab initio} simulation package (VASP) and Phonopy. To overcome the calculational challenge posed by infinite terms, the simplification of the remainder term $R_n$ is essential for practical calculations. 

\subsection{Neutron total scattering}

A neutron total scattering experiment is conducted at NOVA, J-PARC using 0.6 cm$^3$ of prepared CuBr$_2$ powder. The objective is to investigate changes in interatomic distances within the nucleus by analyzing the sample at different temperatures. Data within the range of $1\sim 34 \ \text{\AA}^{-1}$ is selected and subjected to transformation for subsequent analysis. 

The PDF $g(r)$ is calculated using the following equation \cite{keen2001comparison,billinge2013towards}:
\begin{equation}\label{gr}
	g(r)=1+\frac{1}{2\pi ^2r\rho}\int_0^\infty \left(\frac{\rm{d}\sigma/\rm{d}\Omega}{N{\left<b\right>}^2}-\frac{\left<b^2\right>}{{\left<b\right>}^2}\right){\rm sin}(Qr)Q{\rm d}Q, 
\end{equation}
in which $\rho$ represents the atomic number density, which is $0.039/\text{\AA}^3$ for CuBr$_2$. $\rm{d}\sigma/\rm{d}\Omega$ corresponds to the differential scattering cross section, $N$ denotes the number of atoms, $b$ stands for the coherent scattering length, and $\left<\cdots\right>$ here indicates the average of all atoms. $g(r)$ provides the distribution of distances between pairs of atoms in the sample. 

\section{Results}

\subsection{Hard X-ray scattering}

In this study, we analyze the intensity of hard X-ray scattering on CuBr$_2$ by combining measurements with calculations. We aim to examine the impact of atomic displacements on the scattering behavior. 

\begin{figure}
	\includegraphics[width=3.2in]{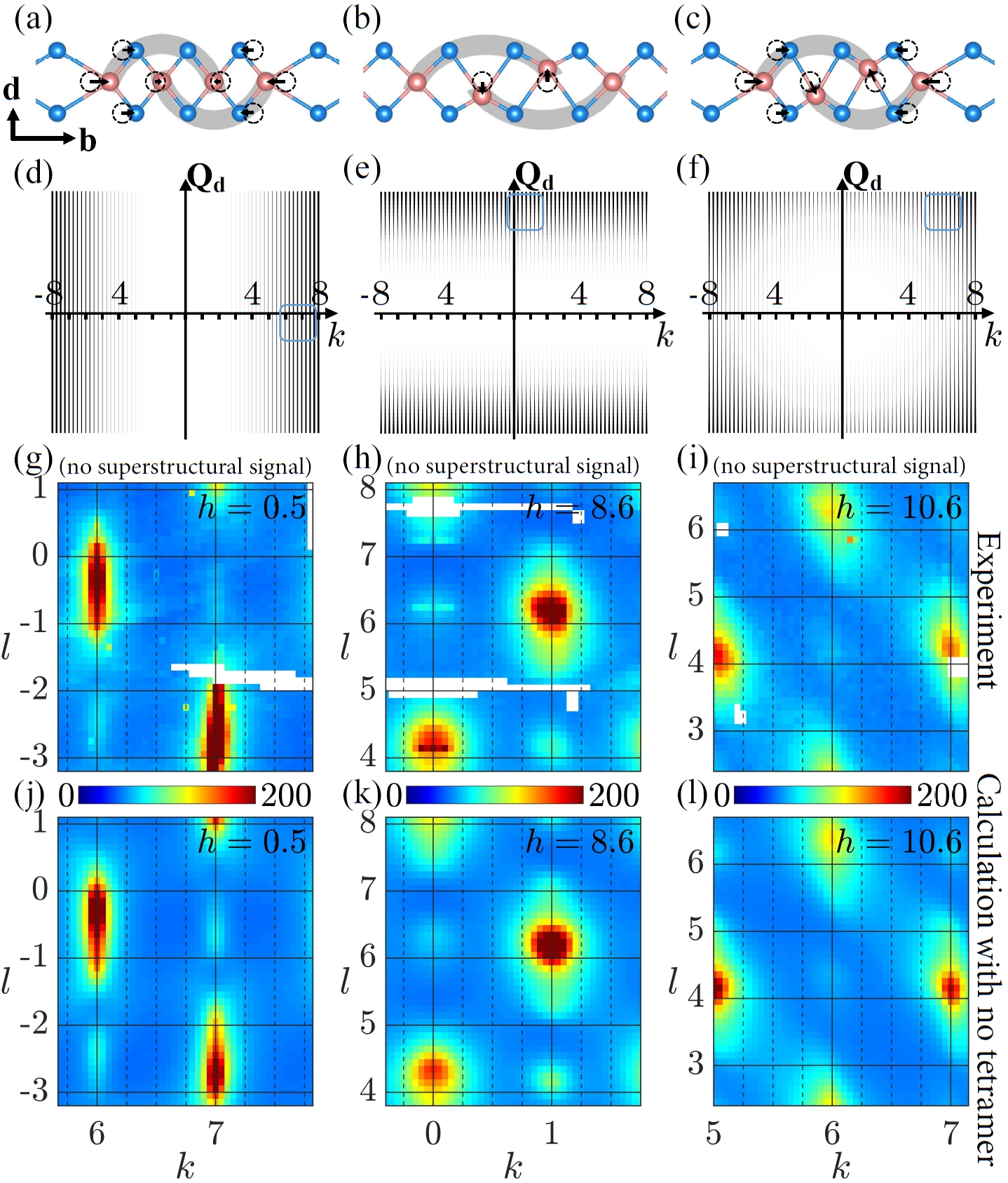}
	\caption{\label{A-L}
		Analysis of the hard X-ray scattering data. (a)-(c) depict three possible type of atomic displacements of tetramers - (a) along $\bf b$, (b) along $\bf d$, and (c) in $\bf b-d$ plane. The dashed circles indicate original atomic positions. (d)-(f) present the estimated distributions of superstructural signals corresponding to (a)-(c). $h$, $k$, $l$ are defined in ${\bf Q}=h{\bf a^*}+k{\bf b^*}+l{\bf c^*}$. ${\bf Q}_{\bf d}$ represents the momentum projection along $\bf d$, approximately parallel to (${\bf a^*}+{\bf c^*}$). The blue rectangles represent the areas of (g)-(i), which respectively depict the experimental intensity at 75 K for large $k$, large ${\bf Q}_{\bf d}$, and both. (j)-(l) present the calculations corresponding to (g)-(i) with no deformations. The intensity unit is $\rm sr^{-1}$. No superstructural signal at $k=\cal{N}/\rm{4}$ is detected anywhere in the experiment. 
	}
\end{figure}

As previously mentioned, tetramerized deformation may result in additional scattering signals occurring at $k = \cal{N}/\rm{4}$. In Fig.~\ref{A-L}(a)-(f), we illustrate three possible types of atomic displacements and the possible superstructural signals of each type. If atoms displace along $\mathbf{b}$, superstructural signals might be present with large $k$. Similarly, if atoms displace along $\mathbf{d}$, superstructural signals might be present with large $\mathbf{Q}_{\mathbf{d}}$. Additionally, for displacements along both $\mathbf{b}$ and $\mathbf{d}$, superstructural signals might be present with large momentum, either in terms of $k$ or $\mathbf{Q}_{\mathbf{d}}$. 

We check the scattering intensity observed with large momentum of $k$, ${\bf Q}_{\bf d}$ or both, as shown in Fig.~\ref{A-L}(g)-(i) respectively. The experimental results do not exhibit any discernible superstructural signals. 

\begin{figure}
	\includegraphics[width=3.2in]{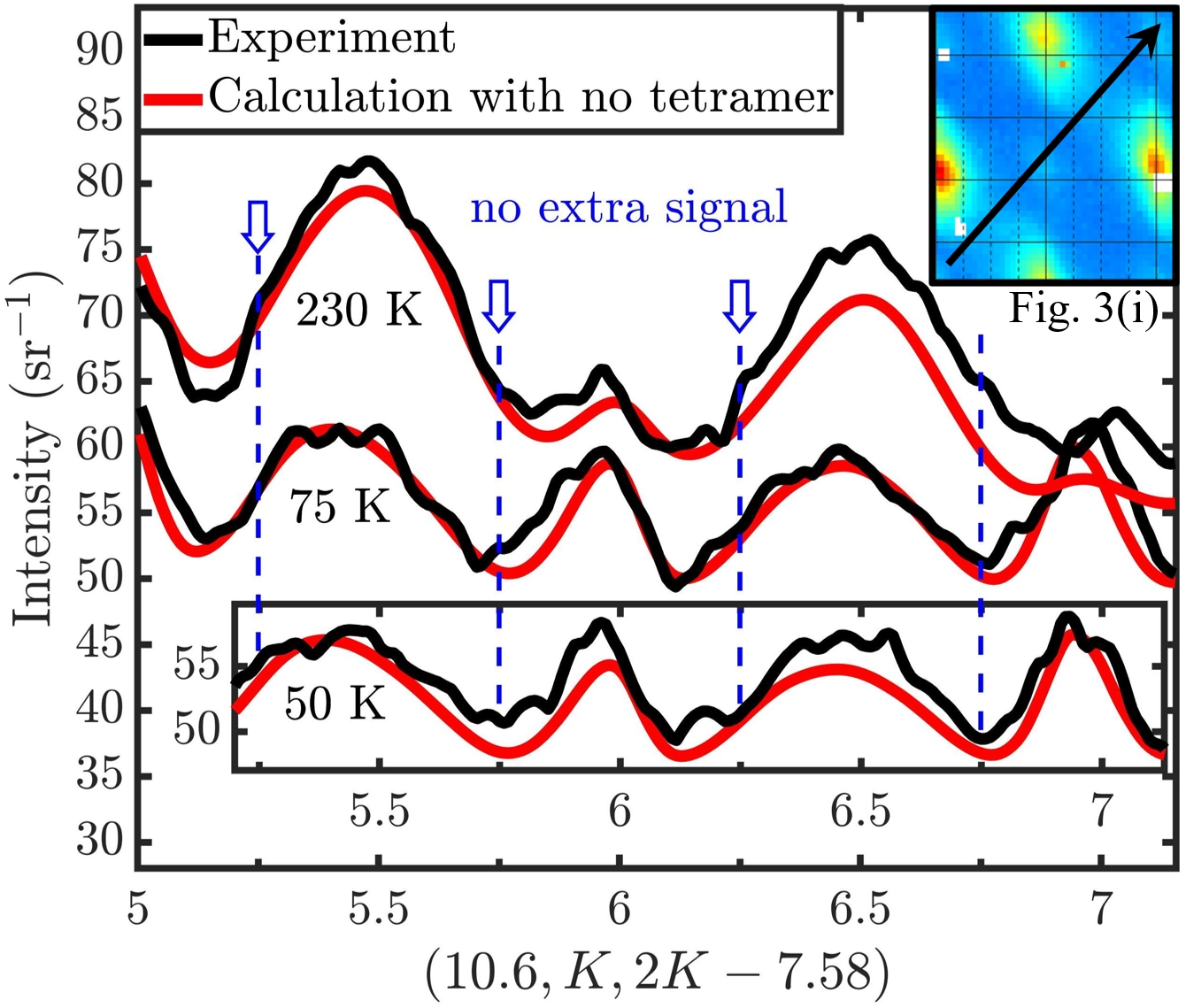}
	\caption{\label{2methods}
		Comparison between experimental and calculated data at various temperatures. The scanning route is the diagonal from the bottom-left to the upper-right of Fig.~\ref{A-L}(i) as shown in the upper-right inset. The experiment agree well with the calculation. No extra signal was detected at $k=\cal{N}/\rm{4}$ in the experimental data. 
	}
\end{figure}

Even so, our investigation uncovers intriguing signals. Unraveling the origin of these signals holds promise in gaining insights into the underlying mechanism governing CuBr$_2$. To explore the complex characteristics of these scattering signals, we conduct calculations of the X-ray scattering intensity including TDS. Fig.~\ref{A-L}(j)-(l) presents the calculated scattering intensity corresponding to the regions depicted in Fig.~\ref{A-L}(g)-(i). Importantly, these calculations are performed on a lattice without tetramerized deformation. Notably, the calculated results faithfully replicate all signals observed in the experimental data, with the exception of minor noise. This comparative analysis between the experiment and calculations indicate the absence of lattice deformations. 

For additional details, we scan along the diagonal of Fig.~\ref{A-L}(i), as shown in Fig.~\ref{2methods}. This examination reveals four distinctive peaks near $K=5.5, 6, 6.5, 7$, each displaying distinct temperature-dependent variations. Upon validation, it becomes evident that the peaks near $K=6, 7$ primarily consist of low-order TDS, while those at $K=5.5, 6.5$ are dominated by high-order TDS. Hence, relying solely on first-order TDS calculations proves insufficient for accurately reproducing the peaks around $K=5.5, 6.5$. Consequently, the calculation of high-order TDS is essential in this study. 

Through the calculations present in this study, we faithfully replicate all the peaks within this dataset, marking a substantial advancement in TDS calculation methodologies. Utilizing the peaks near $K=5.5, 6.5$ as guideposts to map the distribution of high-order TDS signals within the reciprocal space, we trace their origin to the constancy of the nearest Cu-Br distance, confirming the presence of a Cu-Br covalent bond in CuBr$_2$. 

It is worth emphasizing that the calculated data match the experimental data across different temperatures, faithfully reproducing the four peaks and their temperature-dependent behavior observed in the experiment, providing strong evidence of their trivial nature. Importantly, no extra signal is detected at $k=\cal{N}/\rm{4}$. 

We perform a hard X-ray scattering experiment on CuBr$_2$, aligning with TDS calculations excluding tetramers. Results show no evidence of tetramers or spin singlets. 

\subsection{Neutron total scattering}

To detect and analyze the potential tetramerized deformations, we conduct neutron total scattering experiments on CuBr$_2$ and examine the $g(r)$ data presented in Fig.~\ref{gr}. The data reveals multiple distinct peaks, each corresponding to one or several kinds of atomic distance. 

\begin{figure}
	\includegraphics[width=3.2in]{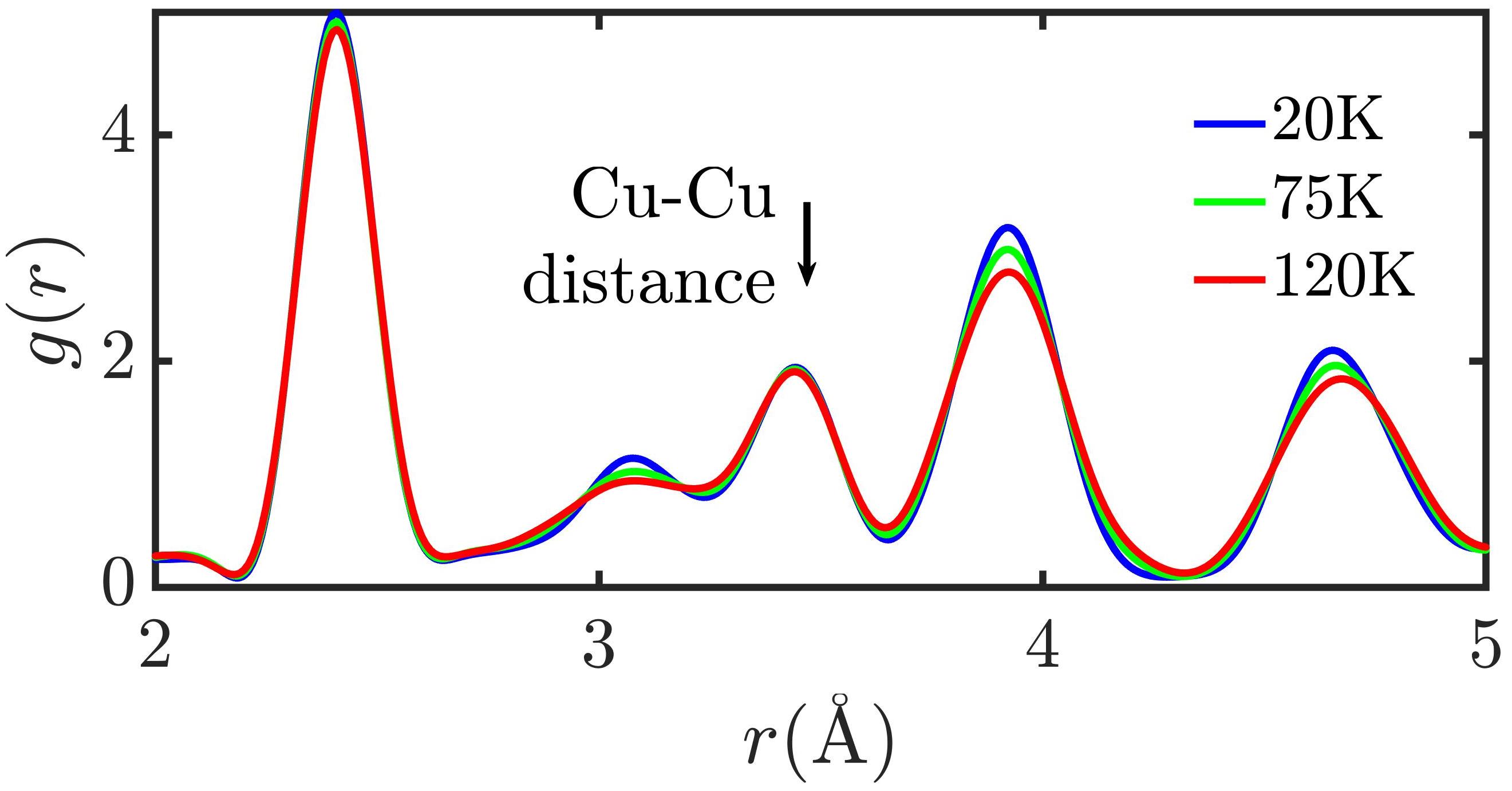}
	\caption{\label{gr}
		The pair distribution function of CuBr$_2$ from the neutron total scattering. No splitting of the Cu-Cu distance is observed. 
	}
\end{figure}

Starting at approximately 2.41 Å, the first peak signifies the nearest Cu-Br bonds. The second peak at around 3.08 Å represents the interchain Cu-Br distance. The integral area of each peak reveals the coordination number. Notably, the second peak varies at higher temperatures due to the weakness of the interchain Cu-Br bond. Weak bonds are unable to endure the thermal vibrations of atoms, resulting in broader peaks at higher temperatures. 

Our focus is on the third peak, which is located at approximately 3.44~Å. This particular peak consists of three different types of atom pairs: the Br-Br pair along $\bf d$, the nearest Br-Br pair along $\bf b$, and the nearest Cu-Cu pairs along $\bf b$. If tetramerized deformation occurs, at least one of the three components in the third peak would split. 

However, upon analyzing Fig.~\ref{gr}, there are no significant temperature-dependent changes observed in the height or width of the third peak. This observation implies the absence of splitting within any of the three components, thereby indicating the lack of significant deformations. 

Our analysis of the neutron total scattering data yields insights, but the characteristics of PDFs do not present any evidence supporting the existence of tetramers or spin singlets within our detection sensitivity. 

\section{Discussion and conclusion}

The scattering experiments have yielded negative results. Despite the absence of superstructural signals in hard X-ray scattering, it is important to explore the potential presence of singlets distributed in a highly random manner. While such a distribution may not give rise to superstructural signals, it could lead to a shorter correlation length and broader Bragg peaks with Lorentzian shapes. It is essential to determine whether the Bragg peaks will indeed exhibit Lorentzian shapes and reach their maximum width in the proximity of $T_{\rm N}$. 

\begin{figure}
	\includegraphics[width=3.2in]{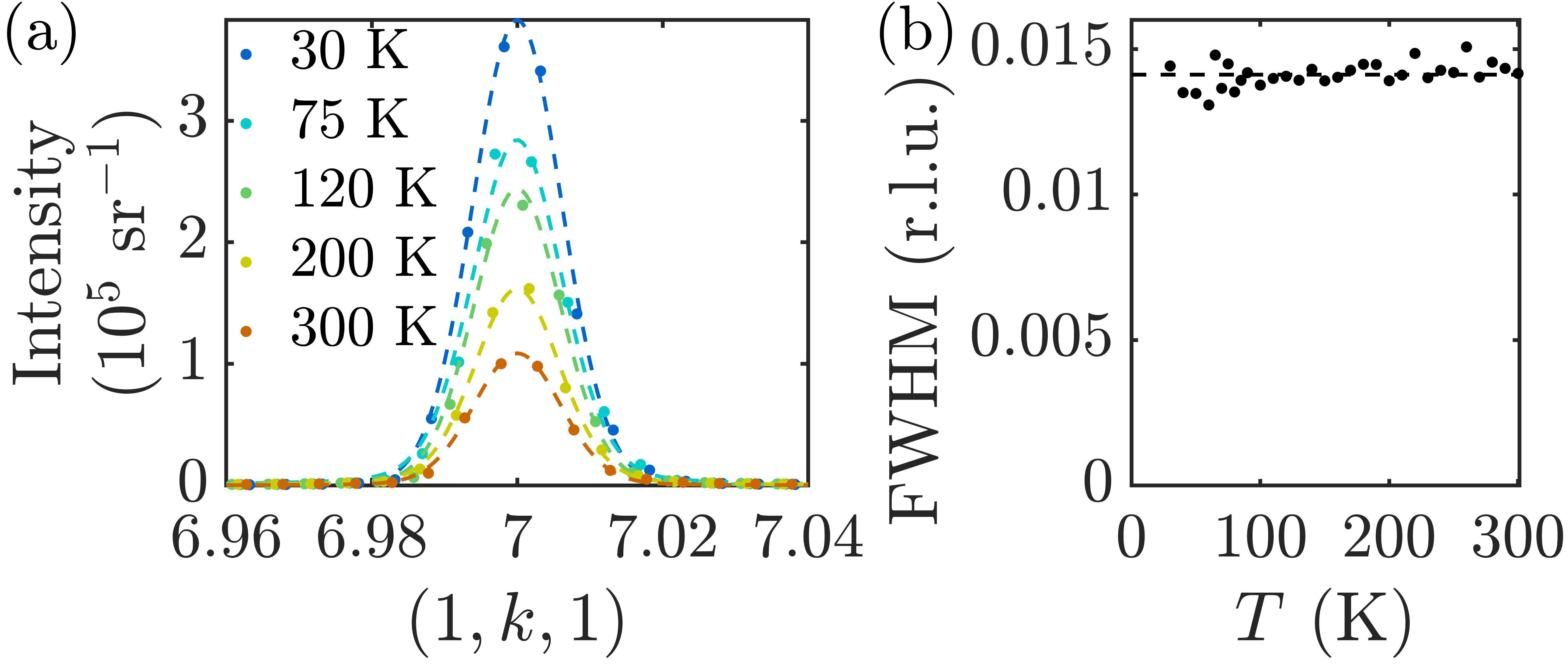}
	\caption{\label{FWHM}
		(a) $k$-scan of the Bragg peak (1,7,1) at various temperatures. The dashed lines represent Gaussian fits. (b) The corresponding full widths at half maximum (FWHM) extracted from the Gaussian fits. The FWHMs remain consistent at various temperatures, with the mean value indicated by the dashed line. These results suggest that any lattice disorder, if present, is minor. 
	}
\end{figure}

We conduct a scan of the (1,7,1) Bragg peak in the hard X-ray data, as depicted in Fig.~\ref{FWHM}. We discover that the peaks could be well-described by Gaussian functions, without any convolution of Lorentzian functions. This finding suggests that the widths of the peaks are primarily influenced by the measurement conditions rather than the changes of correlation length. Thus, any possible lattice disorder, if exists, is below our detection limit. 

The investigation of CuBr$_2$ yielded no evidence of non-magnetic tetramerized order. The underlying mechanism behind the unusual Raman modes in CuBr$_2$ remains to be understood. 

It is supported by multiple studies that magnetoelastic coupling may arise from the hybridization between magnons and phonons \cite{liu2021signature, zhang20203}, which is the other explanation for the unusual Raman modes in CuBr$_2$ \cite{wang2017observation}. The presence of magnon-phonon hybridization may lead to the appearance of Raman peaks in the spectrum, corresponding to the energies of the exciting magnons \cite{wang2017observation}. The temperature dependence of the magnetic order and the thermal excitation of phonons are opposite, which allow the magnon-phonon hybridization exhibits a non-monotonic relationship with temperature. This scenario provides a plausible explanation for the non-monotonic signals in the Raman data of CuBr$_2$. 

In recent years, there has been a growing body of evidence indicating the occurrence of magnon-phonon hybridization in a wide range of low-dimensional materials. Noteworthy instances include FePS$_3$ \cite{liu2021direct, vaclavkova2021magnon}, FePSe$_3$ \cite{luo2023evidence}, and MnPSe$_3$ \cite{mai2021magnon}. These materials have exhibited distinct Raman peaks below their respective $T_{\rm N}$, akin to the partial Raman broad peaks observed in CuBr$_2$ ($e.g.$, the broad peak near 30 meV). CuBr$_2$, being a quasi-one-dimensional antiferromagnet with an effective dimension lower than that of two-dimensional materials. It is reasonable to believe that magnon-phonon hybridization is likely to occur in CuBr$_2$. 

In order to assess the validity of the magnon-phonon hybridization scenario, we carried out experiments encompassing both time-of-flight neutron scatterings and polarized neutron scatterings. The results do not contradict the concept of magnon-phonon hybridization. 

In conclusion, our experimental findings do not substantiate the hypothesis positing the existence of spin singlets in CuBr$_2$, as there is a consistent absence of magnetoelastic deformation across multiple conducted experiments. Instead, our findings suggest that the magnetoelastic coupling in the chain compound CuBr$_2$ is more plausibly ascribed to magnon-phonon hybridization. This conclusion suggests new research paths in phononics and low-dimensional quantum magnetism. 

\begin{acknowledgments}
	
We would like to express our sincere appreciation for the valuable discussions with Hanteng Wang, Hongru Ma, Yiran Liu and Xiquan Zheng. Additionally, we extend our gratitude to the generous support from Junliang Sun's group for providing access to the freeze dryer equipment. The hard X-ray scattering research used resources of the Advanced Photon Source, a U.S. Department of Energy Office of Science User Facility operated for the DOE Office of Science by Argonne National Laboratory under Contract No.~DE-AC02-06CH11357. The neutron experiments at the Materials and Life Science Experimental Facility of the J-PARC are performed under a user program (Proposal No. 2019B0377). We acknowledge financial support by the National Natural Science Foundation of China (Grant No. 11874069). These funding sources have played a essential role in advancing our research efforts.

\end{acknowledgments}

\section*{appendix}
\appendix

\section{Calculation of hard X-ray scattering}

The relationship between the measured hard X-ray scattering intensity and the cross section can be expressed as
\begin{equation}\label{crosec}
	\frac{IL^2}{I_0{\rm cos}^32\theta}=\frac{{\rm d} \sigma}{{\rm d} \it \Omega}=Nr_{\rm e}^2({\rm sin}^2\phi+{\rm cos}^2\phi \ {\rm cos}^22\theta)S, 
\end{equation}
where $I$ is the photon-number illuminance on the detector, $I_0$ is the photon-number flux of the incident X-ray, $L$ is the distance from the sample to the detector plane, $2\theta$ is the scattering angle, $\frac{{\rm d} \sigma}{{\rm d} \it \Omega}$ is the differential scattering cross section, $N$ is the number of the atoms participating in the scattering, $r_{\rm e}$ is the classical electron radius ($r_{\rm e}\approx2.82 \ {\rm fm}$), and $\phi$ is the azimuth angle between the polarization plane of the incident beam and the scattering plane. The data collected directly in the experiment is the photon-number illuminance $I$. Using Eq.~\ref{crosec}, $I$ can be transformed into $S$, whose unit is sr$^{-1}$. The constants of $L$, $I_0$, $r_e$, $etc.$ are inessential. By comparing the experimental $S$ in an arbitrary unit with the calculated $S$ in sr$^{-1}$, the normalization coefficient can be obtained. 

Theoretically, $S$ contains both the coherent scattering $S_{\rm coh}$ and the incoherent scattering ($i.e.$, Compton modified scattering \cite{Warren1990}) $S_{\rm inc}$ that \cite{James1954}
\begin{equation}\label{S=Scoh+Sinc}
	S=S_{\rm coh}+S_{\rm inc}. 
\end{equation}
If all electrons are assumed to be the same, then \cite{James1954}
\begin{equation}\label{Sinc}
	S_{\rm inc}=\frac{1}{\nu}\sum_{d=1}^{\nu}\left(Z_d-\frac{f_d^2}{Z_d}\right), 
\end{equation}
where $Z_d$ is the total number of electrons around an atom (or ion) and $Z_d=f_d(Q=0)$, that $Z_{\rm Cu^{2+}}=27$ and $Z_{\rm Br^-}=36$. $S_{\rm coh}$ can be expanded as \cite{baron2015introduction}
\begin{equation}\label{Scoh}
	S_{\rm coh}=\sum_{p=0}^{\infty}S_p. 
\end{equation}
From Eq.~\ref{Stds=sumSp}, \ref{S=Scoh+Sinc}, \ref{Scoh}, it can be derived that 
\begin{equation}\label{S=S0+Stds+Sinc}
	S=S_0+S_{\rm TDS}+S_{\rm inc}. 
\end{equation}
$S_0$ is Bragg scattering and only appears at $\Gamma$ points. Following the reference \cite{baron2015introduction}, $S_0$ can be derived as \begin{equation}\label{S0b}
	S_0=\frac{1}{\nu}\left|\sum_{d=1}^{\nu}f_de^{i{\bf Q}\cdot{\bf r}_d}e^{-W_d}\right|^2\frac{(2\pi)^3}{V}\delta({\bf Q}-{\bf G}), 
\end{equation}
where $\bf G$ is the momentum of the closest $\Gamma$ point to $\bf Q$, $V$ is the volume of a primitive cell, and $\delta$ is a 3-dimensional delta function satisfies $\iiint \delta({\bf q}) {\mathrm d}q_x{\mathrm d}q_y{\mathrm d}q_z=1$. 
The shape and width of the $\delta$ function depend on the sample and the instrument. 

Since $S_0$ only appear at $\Gamma$ points, the total diffuse scattering intensity is
\begin{equation}\label{Sdif}
	S_{\rm dif}=S-S_0=S_{\rm TDS}+S_{\rm inc}, 
\end{equation}
where $S_{\rm TDS}$ is an infinite series as Eq.~\ref{Stds=sumSp}. 
With the help of the reference \cite{baron2015introduction}, we derive $S_p$ as 
\begin{equation}\label{Sp}
	S_p =\frac{1}{\nu}\frac{1}{p!}\left<
	\sum_{j_1=1}^{3\nu}\cdots \sum_{j_p=1}^{3\nu}
	%	\sum_{j_1,\cdots,j_p=1}^{3\nu}
	\left|\sum_{d=1}^{\nu}f_de^{i{\bf Q}\cdot{\bf r}_d}e^{-W_d}\prod_{m=1}^pF_{{\bf q}_mj_md}\right|^2\right>, 
\end{equation}
where $m$ is the serial number of the phonon, $\left<\cdots \right>$ here means the average for all the situations that satisfy $\sum_{m=1}^p{\bf q}_m={\bf Q}$, and \begin{equation}\label{}
	F_{{\bf q}jd}
	=\left({\bf Q}\cdot{\bf e}_{{\bf q}jd}\right)\sqrt{\frac{{\hbar \ \rm coth}\frac{\hbar\omega_{{\bf q}j}}{2k_{\rm B}T}}{2M_d\omega_{{\bf q}j}}}, 
\end{equation}
where $\omega_{{\bf q}j}$ is the vibrating frequency, and ${\bf e}_{{\bf q}jd}$ is a $3\times 1$ complex vector that denotes the direction and relative phase of the $j$-th vibration mode for the $d$-th atom within the primitive cell. The displacement from the equilibrium position for the $d$-th atom within the $l$-th primitive cell, ${\bf u}_{ld}$, can be expressed with ${\bf e}_{{\bf q}jd}$ as 
\begin{align}\label{}
	{\bf u}_{ld}
	&=\sum_{{\bf q},j}{\bf a}_{{\bf q}jd}\ e^{i{\bf q}\cdot{\bf R}_l-i\omega_{{\bf q}j}t} \\
	&=\frac{1}{\sqrt{M_d}}\sum_{{\bf q},j}c_{{\bf q}j}{\bf e}_{{\bf q}jd}\ e^{i{\bf q}\cdot{\bf R}_l-i\omega_{{\bf q}j}t}, 
\end{align}
where ${\bf R}_l$ denotes the position of the $l$-th primitive cell, ${\bf a}_{{\bf q}jd}$ is the amplitude vector and $c_{{\bf q}j}=\sqrt{\sum_{d}M_d|{\bf a}_{{\bf q}jd}|^2}$. 

${\bf e}_{{\bf q}jd}$ could be derived from force constant matrices. For the harmonic approximation, the corresponding equation of the $ld$-th atom is \cite{baron2015introduction} 
\begin{equation}\label{}
	M_d{\bf \ddot{u}}_{ld}=\sum_{l'=1}^{[N/\nu]}\sum_{d'=1}^\nu\mathbb{K}_{ll'dd'}{\bf u}_{ld},
\end{equation}
where $\mathbb{K}_{ll'dd'}$ is the $3\times 3$ real-valued inter-atomic force constant matrices, describing the stiffness between the $ld$-th and the $l'd'$-th atom. $[N/\nu]$ denotes the total number of primitive cells. The force constant matrices can be transformed to the dynamical matrices $\mathbb{D}_{\bf q}$ as
\begin{equation}\label{}
	\mathbb{D}_{{\bf q}dd'}=\sum_{l'=1}^{[N/\nu]}\frac{\mathbb{K}_{1l'dd'}}{\sqrt{M_dM_{d'}}}e^{i{\bf q}({\bf R}_{l'}-{\bf R}_1)}, 
\end{equation}
where the $3 \times 3$ matrix $\mathbb{D}_{{\bf q}dd'}$ is a component of the $3\nu \times 3\nu$ matrix $\mathbb{D}_{\bf q}$. Our calculation only considers values of $l'$ up to 39, encompassing the nearest 38 primitive cells ($l'=2\sim 39$) of the central primitive cell ($l=1$). For $l'>39$, we assume $\mathbb{K}_{1l'dd'}$ to be $\bf 0$. For $l'\leq 39$, $\mathbb{K}_{1l'dd'}$ is calculated through VASP and Phonopy. For a vibrating mode, the frequency $\omega_{{\bf q}j}$ and the vibrating directions can be obtained from the eigenvalue and the eigenvector of $\mathbb{D}_{\bf q}$ that \cite{baron2015introduction}
\begin{equation}\label{}
	\omega_{{\bf q}j}^2{\bf e}_{{\bf q}j}=\mathbb{D}_{\bf q}{\bf e}_{{\bf q}j}, 
\end{equation}
where the $3\nu\times 1$ complex-valued vector ${\bf e}_{{\bf q}j}$ is made up of ${\bf e}_{{\bf q}jd}$ and $|{\bf e}_{{\bf q}j}|^2=\Sigma _d|{\bf e}_{{\bf q}jd}|^2=1$. 

With $F_{{\bf q}jd}$, $W_d$ can be expressed as
\begin{equation}\label{DWF}
	W_d=\frac{1}{2}\left<\sum_{j=1}^{3\nu}\left|F_{{\bf q}jd}\right|^2\right>_{\bf q},
	%	\propto Q^2{\rm coth}\frac{\Theta _d}{2T},
\end{equation}
where $\left< \cdots \right>_{\bf q}$ means the average for all the ${\bf q}$ in the first Brillouin zone. 
As shown in Eq.~\ref{Stds=sumSp}, there are infinite terms of $S_p$ for $S_{\rm TDS}$, so the remainder term $R_n$ is needed. 
From Eq.~\ref{Sp}, \ref{DWF}, the approximate value of $S_p$ can be derived as
\begin{equation}\label{Spgas}
	S_p\approx\frac{1}{\nu}\sum_{d=1}^{\nu}f_d^2e^{-2W_d}\frac{(2W_d)^p}{p!}. 
\end{equation}

The approximate remainder term can be derived as
\begin{equation}\label{Rngas}
	R_n=\sum_{p=n+1}^{\infty}S_p\approx\frac{1}{\nu}\sum_{d=1}^{\nu}f_d^2e^{-2W_d}\sum_{p=n+1}^{\infty}\frac{(2W_d)^p}{p!}, 
\end{equation}
which can be derived to Eq.~\ref{Rn}. By applying Eq.~\ref{Stds=sumSp} utilizing the expression for $R_n$ as given in Eq.~\ref{Rn}, the value of $S_{\rm TDS}$ can be readily obtained. 

Subsequently, the diffuse scattering $S_{\rm dif}$ can be derived form Eq.~\ref{Sdif}, Eq.~\ref{Stds=sumSp}, Eq.~\ref{Rn}, and Eq.~\ref{Sinc}. Notably, when $n=0$, $S_{\rm dif}$ can be calculated as
\begin{equation}\label{Sdif_}
	S_{\rm dif}^{(n=0)}=\frac{1}{\nu}\sum_{d=1}^{\nu}\left[f_d^2\left(1-e^{-2W_d}\right)+Z_d-\frac{f_d^2}{Z_d}\right], 
\end{equation}
which is the reported approximate formula for diffuse scattering at high temperatures \cite{harvey1933diffuse}. The result is more accurate with larger $n$. In this research, we successfully calculate with $n=5$. 

\normalem % avoid underline
\bibliography{ref}% Produces the bibliography via BibTeX.

\begin{thebibliography}{41}%
\makeatletter
\providecommand \@ifxundefined [1]{%
 \@ifx{#1\undefined}
}%
\providecommand \@ifnum [1]{%
 \ifnum #1\expandafter \@firstoftwo
 \else \expandafter \@secondoftwo
 \fi
}%
\providecommand \@ifx [1]{%
 \ifx #1\expandafter \@firstoftwo
 \else \expandafter \@secondoftwo
 \fi
}%
\providecommand \natexlab [1]{#1}%
\providecommand \enquote  [1]{``#1''}%
\providecommand \bibnamefont  [1]{#1}%
\providecommand \bibfnamefont [1]{#1}%
\providecommand \citenamefont [1]{#1}%
\providecommand \href@noop [0]{\@secondoftwo}%
\providecommand \href [0]{\begingroup \@sanitize@url \@href}%
\providecommand \@href[1]{\@@startlink{#1}\@@href}%
\providecommand \@@href[1]{\endgroup#1\@@endlink}%
\providecommand \@sanitize@url [0]{\catcode `\\12\catcode `\$12\catcode
  `\&12\catcode `\#12\catcode `\^12\catcode `\_12\catcode `\%12\relax}%
\providecommand \@@startlink[1]{}%
\providecommand \@@endlink[0]{}%
\providecommand \url  [0]{\begingroup\@sanitize@url \@url }%
\providecommand \@url [1]{\endgroup\@href {#1}{\urlprefix }}%
\providecommand \urlprefix  [0]{URL }%
\providecommand \Eprint [0]{\href }%
\providecommand \doibase [0]{https://doi.org/}%
\providecommand \selectlanguage [0]{\@gobble}%
\providecommand \bibinfo  [0]{\@secondoftwo}%
\providecommand \bibfield  [0]{\@secondoftwo}%
\providecommand \translation [1]{[#1]}%
\providecommand \BibitemOpen [0]{}%
\providecommand \bibitemStop [0]{}%
\providecommand \bibitemNoStop [0]{.\EOS\space}%
\providecommand \EOS [0]{\spacefactor3000\relax}%
\providecommand \BibitemShut  [1]{\csname bibitem#1\endcsname}%
\let\auto@bib@innerbib\@empty
%</preamble>
\bibitem [{\citenamefont {Hess}(2007)}]{hess2007heat}%
  \BibitemOpen
  \bibfield  {author} {\bibinfo {author} {\bibfnamefont {C.}~\bibnamefont
  {Hess}},\ }\bibfield  {title} {\bibinfo {title} {Heat conduction in
  low-dimensional quantum magnets},\ }\href
  {https://doi.org/10.1140/epjst/e2007-00363-8} {\bibfield  {journal} {\bibinfo
   {journal} {The European Physical Journal Special Topics}\ }\textbf {\bibinfo
  {volume} {151}},\ \bibinfo {pages} {73} (\bibinfo {year} {2007})}\BibitemShut
  {NoStop}%
\bibitem [{\citenamefont {Lemmens}\ \emph {et~al.}(2003)\citenamefont
  {Lemmens}, \citenamefont {G{\"u}ntherodt},\ and\ \citenamefont
  {Gros}}]{lemmens2003magnetic}%
  \BibitemOpen
  \bibfield  {author} {\bibinfo {author} {\bibfnamefont {P.}~\bibnamefont
  {Lemmens}}, \bibinfo {author} {\bibfnamefont {G.}~\bibnamefont
  {G{\"u}ntherodt}},\ and\ \bibinfo {author} {\bibfnamefont {C.}~\bibnamefont
  {Gros}},\ }\bibfield  {title} {\bibinfo {title} {Magnetic light scattering in
  low-dimensional quantum spin systems},\ }\href
  {https://doi.org/10.1016/S0370-1573(02)00321-6} {\bibfield  {journal}
  {\bibinfo  {journal} {Physics Reports}\ }\textbf {\bibinfo {volume} {375}},\
  \bibinfo {pages} {1} (\bibinfo {year} {2003})}\BibitemShut {NoStop}%
\bibitem [{\citenamefont {Sheka}\ \emph {et~al.}(2015)\citenamefont {Sheka},
  \citenamefont {Kravchuk},\ and\ \citenamefont
  {Gaididei}}]{sheka2015curvature}%
  \BibitemOpen
  \bibfield  {author} {\bibinfo {author} {\bibfnamefont {D.~D.}\ \bibnamefont
  {Sheka}}, \bibinfo {author} {\bibfnamefont {V.~P.}\ \bibnamefont
  {Kravchuk}},\ and\ \bibinfo {author} {\bibfnamefont {Y.}~\bibnamefont
  {Gaididei}},\ }\bibfield  {title} {\bibinfo {title} {Curvature effects in
  statics and dynamics of low dimensional magnets},\ }\href
  {https://doi.org/10.1088/1751-8113/48/12/125202} {\bibfield  {journal}
  {\bibinfo  {journal} {Journal of Physics A: Mathematical and Theoretical}\
  }\textbf {\bibinfo {volume} {48}},\ \bibinfo {pages} {125202} (\bibinfo
  {year} {2015})}\BibitemShut {NoStop}%
\bibitem [{\citenamefont {Vasiliev}\ \emph {et~al.}(2018)\citenamefont
  {Vasiliev}, \citenamefont {Volkova}, \citenamefont {Zvereva},\ and\
  \citenamefont {Markina}}]{vasiliev2018milestones}%
  \BibitemOpen
  \bibfield  {author} {\bibinfo {author} {\bibfnamefont {A.}~\bibnamefont
  {Vasiliev}}, \bibinfo {author} {\bibfnamefont {O.}~\bibnamefont {Volkova}},
  \bibinfo {author} {\bibfnamefont {E.}~\bibnamefont {Zvereva}},\ and\ \bibinfo
  {author} {\bibfnamefont {M.}~\bibnamefont {Markina}},\ }\bibfield  {title}
  {\bibinfo {title} {Milestones of low-{D} quantum magnetism},\ }\href
  {https://doi.org/10.1038/s41535-018-0090-7} {\bibfield  {journal} {\bibinfo
  {journal} {npj Quantum Materials}\ }\textbf {\bibinfo {volume} {3}},\
  \bibinfo {pages} {18} (\bibinfo {year} {2018})}\BibitemShut {NoStop}%
\bibitem [{\citenamefont {Bethe}(1931)}]{bethe1931theorie}%
  \BibitemOpen
  \bibfield  {author} {\bibinfo {author} {\bibfnamefont {H.}~\bibnamefont
  {Bethe}},\ }\bibfield  {title} {\bibinfo {title} {Zur theorie der metalle},\
  }\href {https://doi.org/10.1007/BF01341708} {\bibfield  {journal} {\bibinfo
  {journal} {Zeitschrift f{\"u}r Physik}\ }\textbf {\bibinfo {volume} {71}},\
  \bibinfo {pages} {205} (\bibinfo {year} {1931})}\BibitemShut {NoStop}%
\bibitem [{\citenamefont {Aplesnin}(2000)}]{aplesnin2000static}%
  \BibitemOpen
  \bibfield  {author} {\bibinfo {author} {\bibfnamefont {S.}~\bibnamefont
  {Aplesnin}},\ }\bibfield  {title} {\bibinfo {title} {Static and dynamic
  magnetic properties of coupled spin-1/2 antiferromagnetic chains},\ }\href
  {https://doi.org/10.1088/0953-8984/12/37/316} {\bibfield  {journal} {\bibinfo
   {journal} {Journal of Physics: Condensed Matter}\ }\textbf {\bibinfo
  {volume} {12}},\ \bibinfo {pages} {8191} (\bibinfo {year}
  {2000})}\BibitemShut {NoStop}%
\bibitem [{\citenamefont {Kokalj}\ \emph {et~al.}(2015)\citenamefont {Kokalj},
  \citenamefont {Herbrych}, \citenamefont {Zheludev},\ and\ \citenamefont
  {Prelov{\v{s}}ek}}]{kokalj2015antiferromagnetic}%
  \BibitemOpen
  \bibfield  {author} {\bibinfo {author} {\bibfnamefont {J.}~\bibnamefont
  {Kokalj}}, \bibinfo {author} {\bibfnamefont {J.}~\bibnamefont {Herbrych}},
  \bibinfo {author} {\bibfnamefont {A.}~\bibnamefont {Zheludev}},\ and\
  \bibinfo {author} {\bibfnamefont {P.}~\bibnamefont {Prelov{\v{s}}ek}},\
  }\bibfield  {title} {\bibinfo {title} {Antiferromagnetic order in weakly
  coupled random spin chains},\ }\href
  {https://doi.org/10.1103/physrevb.91.155147} {\bibfield  {journal} {\bibinfo
  {journal} {Physical Review B}\ }\textbf {\bibinfo {volume} {91}},\ \bibinfo
  {pages} {155147} (\bibinfo {year} {2015})}\BibitemShut {NoStop}%
\bibitem [{\citenamefont {Shu}\ \emph {et~al.}(2018)\citenamefont {Shu},
  \citenamefont {Dupont}, \citenamefont {Yao}, \citenamefont {Capponi},\ and\
  \citenamefont {Sandvik}}]{shu2018dynamical}%
  \BibitemOpen
  \bibfield  {author} {\bibinfo {author} {\bibfnamefont {Y.-R.}\ \bibnamefont
  {Shu}}, \bibinfo {author} {\bibfnamefont {M.}~\bibnamefont {Dupont}},
  \bibinfo {author} {\bibfnamefont {D.-X.}\ \bibnamefont {Yao}}, \bibinfo
  {author} {\bibfnamefont {S.}~\bibnamefont {Capponi}},\ and\ \bibinfo {author}
  {\bibfnamefont {A.~W.}\ \bibnamefont {Sandvik}},\ }\bibfield  {title}
  {\bibinfo {title} {Dynamical properties of the {$S=\frac{1}{2}$} random
  {Heisenberg} chain},\ }\href {https://doi.org/10.1103/PhysRevB.97.104424}
  {\bibfield  {journal} {\bibinfo  {journal} {Physical Review B}\ }\textbf
  {\bibinfo {volume} {97}},\ \bibinfo {pages} {104424} (\bibinfo {year}
  {2018})}\BibitemShut {NoStop}%
\bibitem [{\citenamefont {Sahling}\ \emph {et~al.}(2015)\citenamefont
  {Sahling}, \citenamefont {Remenyi}, \citenamefont {Paulsen}, \citenamefont
  {Monceau}, \citenamefont {Saligrama}, \citenamefont {Marin}, \citenamefont
  {Revcolevschi}, \citenamefont {Regnault}, \citenamefont {Raymond},\ and\
  \citenamefont {Lorenzo}}]{sahling2015experimental}%
  \BibitemOpen
  \bibfield  {author} {\bibinfo {author} {\bibfnamefont {S.}~\bibnamefont
  {Sahling}}, \bibinfo {author} {\bibfnamefont {G.}~\bibnamefont {Remenyi}},
  \bibinfo {author} {\bibfnamefont {C.}~\bibnamefont {Paulsen}}, \bibinfo
  {author} {\bibfnamefont {P.}~\bibnamefont {Monceau}}, \bibinfo {author}
  {\bibfnamefont {V.}~\bibnamefont {Saligrama}}, \bibinfo {author}
  {\bibfnamefont {C.}~\bibnamefont {Marin}}, \bibinfo {author} {\bibfnamefont
  {A.}~\bibnamefont {Revcolevschi}}, \bibinfo {author} {\bibfnamefont
  {L.}~\bibnamefont {Regnault}}, \bibinfo {author} {\bibfnamefont
  {S.}~\bibnamefont {Raymond}},\ and\ \bibinfo {author} {\bibfnamefont
  {J.}~\bibnamefont {Lorenzo}},\ }\bibfield  {title} {\bibinfo {title}
  {Experimental realization of long-distance entanglement between spins in
  antiferromagnetic quantum spin chains},\ }\href
  {https://doi.org/10.1038/nphys3186} {\bibfield  {journal} {\bibinfo
  {journal} {Nature Physics}\ }\textbf {\bibinfo {volume} {11}},\ \bibinfo
  {pages} {255} (\bibinfo {year} {2015})}\BibitemShut {NoStop}%
\bibitem [{\citenamefont {Haldane}(1982)}]{haldane1982spontaneous}%
  \BibitemOpen
  \bibfield  {author} {\bibinfo {author} {\bibfnamefont {F.}~\bibnamefont
  {Haldane}},\ }\bibfield  {title} {\bibinfo {title} {Spontaneous dimerization
  in the {$S=\frac{1}{2}$} heisenberg antiferromagnetic chain with competing
  interactions},\ }\href {https://doi.org/10.1103/PhysRevB.25.4925} {\bibfield
  {journal} {\bibinfo  {journal} {Physical Review B}\ }\textbf {\bibinfo
  {volume} {25}},\ \bibinfo {pages} {4925} (\bibinfo {year}
  {1982})}\BibitemShut {NoStop}%
\bibitem [{\citenamefont {Wellein}\ \emph {et~al.}(1998)\citenamefont
  {Wellein}, \citenamefont {Fehske},\ and\ \citenamefont
  {Kampf}}]{wellein1998peierls}%
  \BibitemOpen
  \bibfield  {author} {\bibinfo {author} {\bibfnamefont {G.}~\bibnamefont
  {Wellein}}, \bibinfo {author} {\bibfnamefont {H.}~\bibnamefont {Fehske}},\
  and\ \bibinfo {author} {\bibfnamefont {A.~P.}\ \bibnamefont {Kampf}},\
  }\bibfield  {title} {\bibinfo {title} {Peierls dimerization with nonadiabatic
  spin-phonon coupling},\ }\href {https://doi.org/10.1103/physrevlett.81.3956}
  {\bibfield  {journal} {\bibinfo  {journal} {Physical Review Letters}\
  }\textbf {\bibinfo {volume} {81}},\ \bibinfo {pages} {3956} (\bibinfo {year}
  {1998})}\BibitemShut {NoStop}%
\bibitem [{\citenamefont {Becca}\ \emph {et~al.}(2003)\citenamefont {Becca},
  \citenamefont {Mila},\ and\ \citenamefont
  {Poilblanc}}]{becca2003tetramerization}%
  \BibitemOpen
  \bibfield  {author} {\bibinfo {author} {\bibfnamefont {F.}~\bibnamefont
  {Becca}}, \bibinfo {author} {\bibfnamefont {F.}~\bibnamefont {Mila}},\ and\
  \bibinfo {author} {\bibfnamefont {D.}~\bibnamefont {Poilblanc}},\ }\bibfield
  {title} {\bibinfo {title} {Tetramerization of a frustrated spin-1/2 chain},\
  }\href {https://doi.org/10.1103/PhysRevLett.91.067202} {\bibfield  {journal}
  {\bibinfo  {journal} {Physical Review Letters}\ }\textbf {\bibinfo {volume}
  {91}},\ \bibinfo {pages} {067202} (\bibinfo {year} {2003})}\BibitemShut
  {NoStop}%
\bibitem [{\citenamefont {Nishimori}\ \emph {et~al.}(1987)\citenamefont
  {Nishimori}, \citenamefont {Okamoto},\ and\ \citenamefont
  {Yokozawa}}]{nishimori1987competition}%
  \BibitemOpen
  \bibfield  {author} {\bibinfo {author} {\bibfnamefont {H.}~\bibnamefont
  {Nishimori}}, \bibinfo {author} {\bibfnamefont {K.}~\bibnamefont {Okamoto}},\
  and\ \bibinfo {author} {\bibfnamefont {M.}~\bibnamefont {Yokozawa}},\
  }\bibfield  {title} {\bibinfo {title} {Competition between the {N\'eel} and
  the effective singlet states in spin-1/2 alternating {Heisenberg-Ising}
  antiferromagnet in one dimension},\ }\href
  {https://doi.org/10.1143/JPSJ.56.4126} {\bibfield  {journal} {\bibinfo
  {journal} {Journal of the Physical Society of Japan}\ }\textbf {\bibinfo
  {volume} {56}},\ \bibinfo {pages} {4126} (\bibinfo {year}
  {1987})}\BibitemShut {NoStop}%
\bibitem [{\citenamefont {Yamase}\ and\ \citenamefont
  {Kohno}(2004)}]{yamase2004competition}%
  \BibitemOpen
  \bibfield  {author} {\bibinfo {author} {\bibfnamefont {H.}~\bibnamefont
  {Yamase}}\ and\ \bibinfo {author} {\bibfnamefont {H.}~\bibnamefont {Kohno}},\
  }\bibfield  {title} {\bibinfo {title} {Competition and coexistence between
  {Néel} order and d-wave singlet {RVB}},\ }\href
  {https://doi.org/10.1016/j.physc.2004.03.009} {\bibfield  {journal} {\bibinfo
   {journal} {Physica C: Superconductivity}\ }\textbf {\bibinfo {volume}
  {408-410}},\ \bibinfo {pages} {273} (\bibinfo {year} {2004})}\BibitemShut
  {NoStop}%
\bibitem [{\citenamefont {Anderson}(1987)}]{anderson1987resonating}%
  \BibitemOpen
  \bibfield  {author} {\bibinfo {author} {\bibfnamefont {P.~W.}\ \bibnamefont
  {Anderson}},\ }\bibfield  {title} {\bibinfo {title} {The resonating valence
  bond state in {La$_2$CuO$_4$} and superconductivity},\ }\href
  {https://doi.org/10.2307/1698247} {\bibfield  {journal} {\bibinfo  {journal}
  {Science}\ }\textbf {\bibinfo {volume} {235}},\ \bibinfo {pages} {1196}
  (\bibinfo {year} {1987})}\BibitemShut {NoStop}%
\bibitem [{\citenamefont {Anderson}\ \emph {et~al.}(1987)\citenamefont
  {Anderson}, \citenamefont {Baskaran}, \citenamefont {Zou},\ and\
  \citenamefont {Hsu}}]{anderson1987resonatingPRL}%
  \BibitemOpen
  \bibfield  {author} {\bibinfo {author} {\bibfnamefont {P.~W.}\ \bibnamefont
  {Anderson}}, \bibinfo {author} {\bibfnamefont {G.}~\bibnamefont {Baskaran}},
  \bibinfo {author} {\bibfnamefont {Z.}~\bibnamefont {Zou}},\ and\ \bibinfo
  {author} {\bibfnamefont {T.}~\bibnamefont {Hsu}},\ }\bibfield  {title}
  {\bibinfo {title} {Resonating-valence-bond theory of phase transitions and
  superconductivity in {La$_2$CuO$_4$}-based compounds},\ }\href
  {https://doi.org/10.1103/PhysRevLett.58.2790} {\bibfield  {journal} {\bibinfo
   {journal} {Physical Review Letters}\ }\textbf {\bibinfo {volume} {58}},\
  \bibinfo {pages} {2790} (\bibinfo {year} {1987})}\BibitemShut {NoStop}%
\bibitem [{\citenamefont {Anderson}\ \emph {et~al.}(2004)\citenamefont
  {Anderson}, \citenamefont {Lee}, \citenamefont {Randeria}, \citenamefont
  {Rice}, \citenamefont {Trivedi},\ and\ \citenamefont
  {Zhang}}]{anderson2004physics}%
  \BibitemOpen
  \bibfield  {author} {\bibinfo {author} {\bibfnamefont {P.~W.}\ \bibnamefont
  {Anderson}}, \bibinfo {author} {\bibfnamefont {P.}~\bibnamefont {Lee}},
  \bibinfo {author} {\bibfnamefont {M.}~\bibnamefont {Randeria}}, \bibinfo
  {author} {\bibfnamefont {T.}~\bibnamefont {Rice}}, \bibinfo {author}
  {\bibfnamefont {N.}~\bibnamefont {Trivedi}},\ and\ \bibinfo {author}
  {\bibfnamefont {F.}~\bibnamefont {Zhang}},\ }\bibfield  {title} {\bibinfo
  {title} {The physics behind high-temperature superconducting cuprates: the
  ‘plain vanilla’ version of {RVB}},\ }\href
  {https://doi.org/10.1088/0953-8984/16/24/R02} {\bibfield  {journal} {\bibinfo
   {journal} {Journal of Physics: Condensed Matter}\ }\textbf {\bibinfo
  {volume} {16}},\ \bibinfo {pages} {R755} (\bibinfo {year}
  {2004})}\BibitemShut {NoStop}%
\bibitem [{\citenamefont {Ninomiya}\ \emph {et~al.}(2003)\citenamefont
  {Ninomiya}, \citenamefont {Isobe}, \citenamefont {Ueda}, \citenamefont
  {Nishi}, \citenamefont {Ohoyama}, \citenamefont {Sawa},\ and\ \citenamefont
  {Ohama}}]{ninomiya2003observation}%
  \BibitemOpen
  \bibfield  {author} {\bibinfo {author} {\bibfnamefont {E.}~\bibnamefont
  {Ninomiya}}, \bibinfo {author} {\bibfnamefont {M.}~\bibnamefont {Isobe}},
  \bibinfo {author} {\bibfnamefont {Y.}~\bibnamefont {Ueda}}, \bibinfo {author}
  {\bibfnamefont {M.}~\bibnamefont {Nishi}}, \bibinfo {author} {\bibfnamefont
  {K.}~\bibnamefont {Ohoyama}}, \bibinfo {author} {\bibfnamefont
  {H.}~\bibnamefont {Sawa}},\ and\ \bibinfo {author} {\bibfnamefont
  {T.}~\bibnamefont {Ohama}},\ }\bibfield  {title} {\bibinfo {title}
  {Observation of lattice dimerization in spin-singlet low temperature phase of
  {NaTiSi$_2$O$_6$}},\ }\href {https://doi.org/10.1016/S0921-4526(02)02559-0}
  {\bibfield  {journal} {\bibinfo  {journal} {Physica B: Condensed Matter}\
  }\textbf {\bibinfo {volume} {329}},\ \bibinfo {pages} {884} (\bibinfo {year}
  {2003})}\BibitemShut {NoStop}%
\bibitem [{\citenamefont {Deng}\ \emph {et~al.}(2013)\citenamefont {Deng},
  \citenamefont {Kenzelmann}, \citenamefont {Danilkin}, \citenamefont {Studer},
  \citenamefont {Pomjakushin}, \citenamefont {Imperia}, \citenamefont
  {Pomjakushina},\ and\ \citenamefont {Conder}}]{deng2013coexistence}%
  \BibitemOpen
  \bibfield  {author} {\bibinfo {author} {\bibfnamefont {G.}~\bibnamefont
  {Deng}}, \bibinfo {author} {\bibfnamefont {M.}~\bibnamefont {Kenzelmann}},
  \bibinfo {author} {\bibfnamefont {S.}~\bibnamefont {Danilkin}}, \bibinfo
  {author} {\bibfnamefont {A.~J.}\ \bibnamefont {Studer}}, \bibinfo {author}
  {\bibfnamefont {V.}~\bibnamefont {Pomjakushin}}, \bibinfo {author}
  {\bibfnamefont {P.}~\bibnamefont {Imperia}}, \bibinfo {author} {\bibfnamefont
  {E.}~\bibnamefont {Pomjakushina}},\ and\ \bibinfo {author} {\bibfnamefont
  {K.}~\bibnamefont {Conder}},\ }\bibfield  {title} {\bibinfo {title}
  {Coexistence of long-range magnetic ordering and singlet ground state in the
  spin-ladder superconductor {SrCa$_{13}$Cu$_{24}$O$_{41}$}},\ }\href
  {https://doi.org/10.1103/physrevb.88.174424} {\bibfield  {journal} {\bibinfo
  {journal} {Physical Review B}\ }\textbf {\bibinfo {volume} {88}},\ \bibinfo
  {pages} {174424} (\bibinfo {year} {2013})}\BibitemShut {NoStop}%
\bibitem [{\citenamefont {Pasco}\ \emph {et~al.}(2019)\citenamefont {Pasco},
  \citenamefont {El~Baggari}, \citenamefont {Bianco}, \citenamefont
  {Kourkoutis},\ and\ \citenamefont {McQueen}}]{pasco2019tunable}%
  \BibitemOpen
  \bibfield  {author} {\bibinfo {author} {\bibfnamefont {C.~M.}\ \bibnamefont
  {Pasco}}, \bibinfo {author} {\bibfnamefont {I.}~\bibnamefont {El~Baggari}},
  \bibinfo {author} {\bibfnamefont {E.}~\bibnamefont {Bianco}}, \bibinfo
  {author} {\bibfnamefont {L.~F.}\ \bibnamefont {Kourkoutis}},\ and\ \bibinfo
  {author} {\bibfnamefont {T.~M.}\ \bibnamefont {McQueen}},\ }\bibfield
  {title} {\bibinfo {title} {Tunable magnetic transition to a singlet ground
  state in a {2D} van der {Waals} layered trimerized kagom\'e magnet},\ }\href
  {https://doi.org/10.1021/acsnano.9b04392} {\bibfield  {journal} {\bibinfo
  {journal} {ACS nano}\ }\textbf {\bibinfo {volume} {13}},\ \bibinfo {pages}
  {9457} (\bibinfo {year} {2019})}\BibitemShut {NoStop}%
\bibitem [{\citenamefont {Danilovich}\ \emph {et~al.}(2017)\citenamefont
  {Danilovich}, \citenamefont {Karpova}, \citenamefont {Morozov}, \citenamefont
  {Ushakov}, \citenamefont {Streltsov}, \citenamefont {Shakin}, \citenamefont
  {Volkova}, \citenamefont {Zvereva},\ and\ \citenamefont
  {Vasiliev}}]{danilovich2017spin}%
  \BibitemOpen
  \bibfield  {author} {\bibinfo {author} {\bibfnamefont {I.~L.}\ \bibnamefont
  {Danilovich}}, \bibinfo {author} {\bibfnamefont {E.~V.}\ \bibnamefont
  {Karpova}}, \bibinfo {author} {\bibfnamefont {I.~V.}\ \bibnamefont
  {Morozov}}, \bibinfo {author} {\bibfnamefont {A.~V.}\ \bibnamefont
  {Ushakov}}, \bibinfo {author} {\bibfnamefont {S.~V.}\ \bibnamefont
  {Streltsov}}, \bibinfo {author} {\bibfnamefont {A.~A.}\ \bibnamefont
  {Shakin}}, \bibinfo {author} {\bibfnamefont {O.~S.}\ \bibnamefont {Volkova}},
  \bibinfo {author} {\bibfnamefont {E.~A.}\ \bibnamefont {Zvereva}},\ and\
  \bibinfo {author} {\bibfnamefont {A.~N.}\ \bibnamefont {Vasiliev}},\
  }\bibfield  {title} {\bibinfo {title} {Spin-singlet quantum ground state in
  zigzag spin ladder {Cu(CF$_3$COO)$_2$}},\ }\href
  {https://doi.org/10.1002/cphc.201700707} {\bibfield  {journal} {\bibinfo
  {journal} {ChemPhysChem}\ }\textbf {\bibinfo {volume} {18}},\ \bibinfo
  {pages} {2482} (\bibinfo {year} {2017})}\BibitemShut {NoStop}%
\bibitem [{\citenamefont {Wang}\ \emph {et~al.}(2017)\citenamefont {Wang},
  \citenamefont {Yu}, \citenamefont {Liu}, \citenamefont {Chen}, \citenamefont
  {Du}, \citenamefont {Hu}, \citenamefont {Wang}, \citenamefont {Iida},
  \citenamefont {Kamazawa}, \citenamefont {Wakimoto} \emph
  {et~al.}}]{wang2017observation}%
  \BibitemOpen
  \bibfield  {author} {\bibinfo {author} {\bibfnamefont {C.}~\bibnamefont
  {Wang}}, \bibinfo {author} {\bibfnamefont {D.}~\bibnamefont {Yu}}, \bibinfo
  {author} {\bibfnamefont {X.}~\bibnamefont {Liu}}, \bibinfo {author}
  {\bibfnamefont {R.}~\bibnamefont {Chen}}, \bibinfo {author} {\bibfnamefont
  {X.}~\bibnamefont {Du}}, \bibinfo {author} {\bibfnamefont {B.}~\bibnamefont
  {Hu}}, \bibinfo {author} {\bibfnamefont {L.}~\bibnamefont {Wang}}, \bibinfo
  {author} {\bibfnamefont {K.}~\bibnamefont {Iida}}, \bibinfo {author}
  {\bibfnamefont {K.}~\bibnamefont {Kamazawa}}, \bibinfo {author}
  {\bibfnamefont {S.}~\bibnamefont {Wakimoto}}, \emph {et~al.},\ }\bibfield
  {title} {\bibinfo {title} {Observation of magnetoelastic effects in a
  quasi-one-dimensional spiral magnet},\ }\href
  {https://doi.org/10.1103/PhysRevB.96.085111} {\bibfield  {journal} {\bibinfo
  {journal} {Physical Review B}\ }\textbf {\bibinfo {volume} {96}},\ \bibinfo
  {pages} {085111} (\bibinfo {year} {2017})}\BibitemShut {NoStop}%
\bibitem [{\citenamefont {Zhao}\ \emph {et~al.}(2012)\citenamefont {Zhao},
  \citenamefont {Hung}, \citenamefont {Li}, \citenamefont {Chen}, \citenamefont
  {Wu}, \citenamefont {Kremer}, \citenamefont {Banks}, \citenamefont {Simon},
  \citenamefont {Whangbo}, \citenamefont {Lee} \emph {et~al.}}]{zhao2012cubr2}%
  \BibitemOpen
  \bibfield  {author} {\bibinfo {author} {\bibfnamefont {L.}~\bibnamefont
  {Zhao}}, \bibinfo {author} {\bibfnamefont {T.-L.}\ \bibnamefont {Hung}},
  \bibinfo {author} {\bibfnamefont {C.-C.}\ \bibnamefont {Li}}, \bibinfo
  {author} {\bibfnamefont {Y.-Y.}\ \bibnamefont {Chen}}, \bibinfo {author}
  {\bibfnamefont {M.-K.}\ \bibnamefont {Wu}}, \bibinfo {author} {\bibfnamefont
  {R.~K.}\ \bibnamefont {Kremer}}, \bibinfo {author} {\bibfnamefont {M.~G.}\
  \bibnamefont {Banks}}, \bibinfo {author} {\bibfnamefont {A.}~\bibnamefont
  {Simon}}, \bibinfo {author} {\bibfnamefont {M.-H.}\ \bibnamefont {Whangbo}},
  \bibinfo {author} {\bibfnamefont {C.}~\bibnamefont {Lee}}, \emph {et~al.},\
  }\bibfield  {title} {\bibinfo {title} {{CuBr$_2$} -- a new multiferroic
  material with high critical temperature},\ }\href
  {https://doi.org/10.1002/adma.201200734} {\bibfield  {journal} {\bibinfo
  {journal} {Advanced Materials}\ }\textbf {\bibinfo {volume} {24}},\ \bibinfo
  {pages} {2469} (\bibinfo {year} {2012})}\BibitemShut {NoStop}%
\bibitem [{\citenamefont {Wang}\ \emph {et~al.}(2018)\citenamefont {Wang},
  \citenamefont {Zheng}, \citenamefont {Chen}, \citenamefont {Wang},
  \citenamefont {Zhang}, \citenamefont {Cui}, \citenamefont {Wang},
  \citenamefont {Li}, \citenamefont {Xu}, \citenamefont {Yuan} \emph
  {et~al.}}]{wang2018nmr}%
  \BibitemOpen
  \bibfield  {author} {\bibinfo {author} {\bibfnamefont {R.-Q.}\ \bibnamefont
  {Wang}}, \bibinfo {author} {\bibfnamefont {J.-C.}\ \bibnamefont {Zheng}},
  \bibinfo {author} {\bibfnamefont {T.}~\bibnamefont {Chen}}, \bibinfo {author}
  {\bibfnamefont {P.-S.}\ \bibnamefont {Wang}}, \bibinfo {author}
  {\bibfnamefont {J.-S.}\ \bibnamefont {Zhang}}, \bibinfo {author}
  {\bibfnamefont {Y.}~\bibnamefont {Cui}}, \bibinfo {author} {\bibfnamefont
  {C.}~\bibnamefont {Wang}}, \bibinfo {author} {\bibfnamefont {Y.}~\bibnamefont
  {Li}}, \bibinfo {author} {\bibfnamefont {S.}~\bibnamefont {Xu}}, \bibinfo
  {author} {\bibfnamefont {F.}~\bibnamefont {Yuan}}, \emph {et~al.},\
  }\bibfield  {title} {\bibinfo {title} {{NMR} evidence of charge fluctuations
  in multiferroic {CuBr$_2$}},\ }\href
  {https://doi.org/10.1088/1674-1056/27/3/037502} {\bibfield  {journal}
  {\bibinfo  {journal} {Chinese Physics B}\ }\textbf {\bibinfo {volume} {27}},\
  \bibinfo {pages} {037502} (\bibinfo {year} {2018})}\BibitemShut {NoStop}%
\bibitem [{\citenamefont {Apostolov}\ \emph {et~al.}(2022)\citenamefont
  {Apostolov}, \citenamefont {Apostolova},\ and\ \citenamefont
  {Wesselinowa}}]{apostolov2022magnetic}%
  \BibitemOpen
  \bibfield  {author} {\bibinfo {author} {\bibfnamefont {A.}~\bibnamefont
  {Apostolov}}, \bibinfo {author} {\bibfnamefont {I.}~\bibnamefont
  {Apostolova}},\ and\ \bibinfo {author} {\bibfnamefont {J.}~\bibnamefont
  {Wesselinowa}},\ }\bibfield  {title} {\bibinfo {title} {Magnetic and electric
  properties of multiferroic {CuBr$_2$}},\ }\href
  {https://doi.org/10.1016/j.jmmm.2022.169633} {\bibfield  {journal} {\bibinfo
  {journal} {Journal of Magnetism and Magnetic Materials}\ }\textbf {\bibinfo
  {volume} {560}},\ \bibinfo {pages} {169633} (\bibinfo {year}
  {2022})}\BibitemShut {NoStop}%
\bibitem [{\citenamefont {Lee}\ \emph {et~al.}(2012)\citenamefont {Lee},
  \citenamefont {Liu}, \citenamefont {Whangbo}, \citenamefont {Koo},
  \citenamefont {Kremer},\ and\ \citenamefont {Simon}}]{lee2012investigation}%
  \BibitemOpen
  \bibfield  {author} {\bibinfo {author} {\bibfnamefont {C.}~\bibnamefont
  {Lee}}, \bibinfo {author} {\bibfnamefont {J.}~\bibnamefont {Liu}}, \bibinfo
  {author} {\bibfnamefont {M.-H.}\ \bibnamefont {Whangbo}}, \bibinfo {author}
  {\bibfnamefont {H.-J.}\ \bibnamefont {Koo}}, \bibinfo {author} {\bibfnamefont
  {R.}~\bibnamefont {Kremer}},\ and\ \bibinfo {author} {\bibfnamefont
  {A.}~\bibnamefont {Simon}},\ }\bibfield  {title} {\bibinfo {title}
  {Investigation of the spin exchange interactions and the magnetic structure
  of the high-temperature multiferroic {CuBr$_2$}},\ }\href
  {https://doi.org/10.1103/PhysRevB.86.060407} {\bibfield  {journal} {\bibinfo
  {journal} {Physical Review B}\ }\textbf {\bibinfo {volume} {86}},\ \bibinfo
  {pages} {060407} (\bibinfo {year} {2012})}\BibitemShut {NoStop}%
\bibitem [{\citenamefont {Lebernegg}\ \emph {et~al.}(2013)\citenamefont
  {Lebernegg}, \citenamefont {Schmitt}, \citenamefont {Tsirlin}, \citenamefont
  {Janson},\ and\ \citenamefont {Rosner}}]{lebernegg2013magnetism}%
  \BibitemOpen
  \bibfield  {author} {\bibinfo {author} {\bibfnamefont {S.}~\bibnamefont
  {Lebernegg}}, \bibinfo {author} {\bibfnamefont {M.}~\bibnamefont {Schmitt}},
  \bibinfo {author} {\bibfnamefont {A.~A.}\ \bibnamefont {Tsirlin}}, \bibinfo
  {author} {\bibfnamefont {O.}~\bibnamefont {Janson}},\ and\ \bibinfo {author}
  {\bibfnamefont {H.}~\bibnamefont {Rosner}},\ }\bibfield  {title} {\bibinfo
  {title} {Magnetism of {CuX$_2$} frustrated chains ({X = F, Cl, Br}): the role
  of covalency},\ }\href {https://doi.org/10.1103/PhysRevB.87.155111}
  {\bibfield  {journal} {\bibinfo  {journal} {Phys. Rev. B}\ }\textbf {\bibinfo
  {volume} {87}},\ \bibinfo {pages} {155111} (\bibinfo {year}
  {2013})}\BibitemShut {NoStop}%
\bibitem [{\citenamefont {Warren}(1969)}]{Warren1990}%
  \BibitemOpen
  \bibfield  {author} {\bibinfo {author} {\bibfnamefont {B.~E.}\ \bibnamefont
  {Warren}},\ }\href {https://archive.org/details/xraydiffraction00warr} {\emph
  {\bibinfo {title} {X-ray diffraction}}}\ (\bibinfo  {publisher}
  {Addison-Wesley Publishing Company},\ \bibinfo {year} {1969})\BibitemShut
  {NoStop}%
\bibitem [{\citenamefont {James}(1962)}]{James1954}%
  \BibitemOpen
  \bibfield  {author} {\bibinfo {author} {\bibfnamefont {R.~W.}\ \bibnamefont
  {James}},\ }\href
  {https://archive.org/details/opticalprinciple031059mbp/page/n7/mode/2up}
  {\emph {\bibinfo {title} {The optical principles of the diffraction of
  X-rays}}}\ (\bibinfo  {publisher} {George Bell \& Sons},\ \bibinfo {year}
  {1962})\BibitemShut {NoStop}%
\bibitem [{\citenamefont {Als-Nielsen}\ and\ \citenamefont
  {McMorrow}(2011)}]{AlsNielsen2011elements}%
  \BibitemOpen
  \bibfield  {author} {\bibinfo {author} {\bibfnamefont {J.}~\bibnamefont
  {Als-Nielsen}}\ and\ \bibinfo {author} {\bibfnamefont {D.}~\bibnamefont
  {McMorrow}},\ }\href {https://doi.org/10.1002/9781119998365} {\emph {\bibinfo
  {title} {Elements of modern X-ray physics}}}\ (\bibinfo  {publisher} {John
  Wiley \& Sons},\ \bibinfo {year} {2011})\BibitemShut {NoStop}%
\bibitem [{\citenamefont {Baron}(2015)}]{baron2015introduction}%
  \BibitemOpen
  \bibfield  {author} {\bibinfo {author} {\bibfnamefont {A.~Q.}\ \bibnamefont
  {Baron}},\ }\bibfield  {title} {\bibinfo {title} {Introduction to
  high-resolution inelastic {X-ray} scattering},\ }\href
  {https://arxiv.org/abs/1504.01098} {\bibfield  {journal} {\bibinfo  {journal}
  {arXiv:1504.01098}\ } (\bibinfo {year} {2015})}\BibitemShut {NoStop}%
\bibitem [{\citenamefont {Brown}\ \emph {et~al.}(2006)\citenamefont {Brown},
  \citenamefont {Fox}, \citenamefont {Maslen}, \citenamefont {O'Keefe},\ and\
  \citenamefont {Willis}}]{brown2006Intensity}%
  \BibitemOpen
  \bibfield  {author} {\bibinfo {author} {\bibfnamefont {P.}~\bibnamefont
  {Brown}}, \bibinfo {author} {\bibfnamefont {A.}~\bibnamefont {Fox}}, \bibinfo
  {author} {\bibfnamefont {E.}~\bibnamefont {Maslen}}, \bibinfo {author}
  {\bibfnamefont {M.}~\bibnamefont {O'Keefe}},\ and\ \bibinfo {author}
  {\bibfnamefont {B.}~\bibnamefont {Willis}},\ }\bibfield  {title} {\bibinfo
  {title} {Intensity of diffracted intensities},\ }\href
  {https://it.iucr.org/Cb/ch6o1v0001/sec6o1o1/} {\bibfield  {journal} {\bibinfo
   {journal} {International Tables for Crystallography}\ }\textbf {\bibinfo
  {volume} {C}},\ \bibinfo {pages} {554} (\bibinfo {year} {2006})}\BibitemShut
  {NoStop}%
\bibitem [{\citenamefont {Keen}(2001)}]{keen2001comparison}%
  \BibitemOpen
  \bibfield  {author} {\bibinfo {author} {\bibfnamefont {D.~A.}\ \bibnamefont
  {Keen}},\ }\bibfield  {title} {\bibinfo {title} {A comparison of various
  commonly used correlation functions for describing total scattering},\ }\href
  {https://doi.org/10.1107/S0021889800019993} {\bibfield  {journal} {\bibinfo
  {journal} {Journal of Applied Crystallography}\ }\textbf {\bibinfo {volume}
  {34}},\ \bibinfo {pages} {172} (\bibinfo {year} {2001})}\BibitemShut
  {NoStop}%
\bibitem [{\citenamefont {Billinge}\ and\ \citenamefont
  {Farrow}(2013)}]{billinge2013towards}%
  \BibitemOpen
  \bibfield  {author} {\bibinfo {author} {\bibfnamefont {S.~J.}\ \bibnamefont
  {Billinge}}\ and\ \bibinfo {author} {\bibfnamefont {C.~L.}\ \bibnamefont
  {Farrow}},\ }\bibfield  {title} {\bibinfo {title} {Towards a robust ad hoc
  data correction approach that yields reliable atomic pair distribution
  functions from powder diffraction data},\ }\href
  {https://doi.org/10.1088/0953-8984/25/45/454202} {\bibfield  {journal}
  {\bibinfo  {journal} {Journal of Physics: Condensed Matter}\ }\textbf
  {\bibinfo {volume} {25}},\ \bibinfo {pages} {454202} (\bibinfo {year}
  {2013})}\BibitemShut {NoStop}%
\bibitem [{\citenamefont {Liu}\ \emph {et~al.}(2021{\natexlab{a}})\citenamefont
  {Liu}, \citenamefont {Xiao}, \citenamefont {Siemann}, \citenamefont {Weber},
  \citenamefont {Andres}, \citenamefont {Bronsch}, \citenamefont {Oppeneer},\
  and\ \citenamefont {Weinelt}}]{liu2021signature}%
  \BibitemOpen
  \bibfield  {author} {\bibinfo {author} {\bibfnamefont {B.}~\bibnamefont
  {Liu}}, \bibinfo {author} {\bibfnamefont {H.}~\bibnamefont {Xiao}}, \bibinfo
  {author} {\bibfnamefont {G.}~\bibnamefont {Siemann}}, \bibinfo {author}
  {\bibfnamefont {J.}~\bibnamefont {Weber}}, \bibinfo {author} {\bibfnamefont
  {B.}~\bibnamefont {Andres}}, \bibinfo {author} {\bibfnamefont
  {W.}~\bibnamefont {Bronsch}}, \bibinfo {author} {\bibfnamefont {P.~M.}\
  \bibnamefont {Oppeneer}},\ and\ \bibinfo {author} {\bibfnamefont
  {M.}~\bibnamefont {Weinelt}},\ }\bibfield  {title} {\bibinfo {title}
  {Signature of magnon polarons in electron relaxation on terbium revealed by
  comparison with gadolinium},\ }\href
  {https://doi.org/10.1103/PhysRevB.104.024434} {\bibfield  {journal} {\bibinfo
   {journal} {Physical Review B}\ }\textbf {\bibinfo {volume} {104}},\ \bibinfo
  {pages} {024434} (\bibinfo {year} {2021}{\natexlab{a}})}\BibitemShut
  {NoStop}%
\bibitem [{\citenamefont {Zhang}\ \emph {et~al.}(2020)\citenamefont {Zhang},
  \citenamefont {Go}, \citenamefont {Lee},\ and\ \citenamefont
  {Kim}}]{zhang20203}%
  \BibitemOpen
  \bibfield  {author} {\bibinfo {author} {\bibfnamefont {S.}~\bibnamefont
  {Zhang}}, \bibinfo {author} {\bibfnamefont {G.}~\bibnamefont {Go}}, \bibinfo
  {author} {\bibfnamefont {K.-J.}\ \bibnamefont {Lee}},\ and\ \bibinfo {author}
  {\bibfnamefont {S.~K.}\ \bibnamefont {Kim}},\ }\bibfield  {title} {\bibinfo
  {title} {{SU(3)} topology of magnon-phonon hybridization in {2D}
  antiferromagnets},\ }\href {https://doi.org/10.1103/PhysRevLett.124.147204}
  {\bibfield  {journal} {\bibinfo  {journal} {Physical Review Letters}\
  }\textbf {\bibinfo {volume} {124}},\ \bibinfo {pages} {147204} (\bibinfo
  {year} {2020})}\BibitemShut {NoStop}%
\bibitem [{\citenamefont {Liu}\ \emph {et~al.}(2021{\natexlab{b}})\citenamefont
  {Liu}, \citenamefont {Del~{\'A}guila}, \citenamefont {Bhowmick},
  \citenamefont {Gan}, \citenamefont {Do}, \citenamefont {Prosnikov},
  \citenamefont {Sedmidubsk{\`y}}, \citenamefont {Sofer}, \citenamefont
  {Christianen}, \citenamefont {Sengupta},\ and\ \citenamefont
  {Xiong}}]{liu2021direct}%
  \BibitemOpen
  \bibfield  {author} {\bibinfo {author} {\bibfnamefont {S.}~\bibnamefont
  {Liu}}, \bibinfo {author} {\bibfnamefont {A.~G.}\ \bibnamefont
  {Del~{\'A}guila}}, \bibinfo {author} {\bibfnamefont {D.}~\bibnamefont
  {Bhowmick}}, \bibinfo {author} {\bibfnamefont {C.~K.}\ \bibnamefont {Gan}},
  \bibinfo {author} {\bibfnamefont {T.~T.~H.}\ \bibnamefont {Do}}, \bibinfo
  {author} {\bibfnamefont {M.}~\bibnamefont {Prosnikov}}, \bibinfo {author}
  {\bibfnamefont {D.}~\bibnamefont {Sedmidubsk{\`y}}}, \bibinfo {author}
  {\bibfnamefont {Z.}~\bibnamefont {Sofer}}, \bibinfo {author} {\bibfnamefont
  {P.~C.}\ \bibnamefont {Christianen}}, \bibinfo {author} {\bibfnamefont
  {P.}~\bibnamefont {Sengupta}},\ and\ \bibinfo {author} {\bibfnamefont
  {Q.}~\bibnamefont {Xiong}},\ }\bibfield  {title} {\bibinfo {title} {Direct
  observation of magnon-phonon strong coupling in two-dimensional
  antiferromagnet at high magnetic fields},\ }\href
  {https://doi.org/10.1103/PhysRevLett.127.097401} {\bibfield  {journal}
  {\bibinfo  {journal} {Physical Review Letters}\ }\textbf {\bibinfo {volume}
  {127}},\ \bibinfo {pages} {097401} (\bibinfo {year}
  {2021}{\natexlab{b}})}\BibitemShut {NoStop}%
\bibitem [{\citenamefont {Vaclavkova}\ \emph {et~al.}(2021)\citenamefont
  {Vaclavkova}, \citenamefont {Palit}, \citenamefont {Wyzula}, \citenamefont
  {Ghosh}, \citenamefont {Delhomme}, \citenamefont {Maity}, \citenamefont
  {Kapuscinski}, \citenamefont {Ghosh}, \citenamefont {Veis}, \citenamefont
  {Grzeszczyk} \emph {et~al.}}]{vaclavkova2021magnon}%
  \BibitemOpen
  \bibfield  {author} {\bibinfo {author} {\bibfnamefont {D.}~\bibnamefont
  {Vaclavkova}}, \bibinfo {author} {\bibfnamefont {M.}~\bibnamefont {Palit}},
  \bibinfo {author} {\bibfnamefont {J.}~\bibnamefont {Wyzula}}, \bibinfo
  {author} {\bibfnamefont {S.}~\bibnamefont {Ghosh}}, \bibinfo {author}
  {\bibfnamefont {A.}~\bibnamefont {Delhomme}}, \bibinfo {author}
  {\bibfnamefont {S.}~\bibnamefont {Maity}}, \bibinfo {author} {\bibfnamefont
  {P.}~\bibnamefont {Kapuscinski}}, \bibinfo {author} {\bibfnamefont
  {A.}~\bibnamefont {Ghosh}}, \bibinfo {author} {\bibfnamefont
  {M.}~\bibnamefont {Veis}}, \bibinfo {author} {\bibfnamefont {M.}~\bibnamefont
  {Grzeszczyk}}, \emph {et~al.},\ }\bibfield  {title} {\bibinfo {title} {Magnon
  polarons in the van der {Waals} antiferromagnet {FePS$_3$}},\ }\href
  {https://doi.org/10.1103/PhysRevB.104.134437} {\bibfield  {journal} {\bibinfo
   {journal} {Physical Review B}\ }\textbf {\bibinfo {volume} {104}},\ \bibinfo
  {pages} {134437} (\bibinfo {year} {2021})}\BibitemShut {NoStop}%
\bibitem [{\citenamefont {Luo}\ \emph {et~al.}(2023)\citenamefont {Luo},
  \citenamefont {Li}, \citenamefont {Ye}, \citenamefont {Xu}, \citenamefont
  {Yan}, \citenamefont {Zhang}, \citenamefont {Ye}, \citenamefont {Chen},
  \citenamefont {Hu}, \citenamefont {Teng} \emph {et~al.}}]{luo2023evidence}%
  \BibitemOpen
  \bibfield  {author} {\bibinfo {author} {\bibfnamefont {J.}~\bibnamefont
  {Luo}}, \bibinfo {author} {\bibfnamefont {S.}~\bibnamefont {Li}}, \bibinfo
  {author} {\bibfnamefont {Z.}~\bibnamefont {Ye}}, \bibinfo {author}
  {\bibfnamefont {R.}~\bibnamefont {Xu}}, \bibinfo {author} {\bibfnamefont
  {H.}~\bibnamefont {Yan}}, \bibinfo {author} {\bibfnamefont {J.}~\bibnamefont
  {Zhang}}, \bibinfo {author} {\bibfnamefont {G.}~\bibnamefont {Ye}}, \bibinfo
  {author} {\bibfnamefont {L.}~\bibnamefont {Chen}}, \bibinfo {author}
  {\bibfnamefont {D.}~\bibnamefont {Hu}}, \bibinfo {author} {\bibfnamefont
  {X.}~\bibnamefont {Teng}}, \emph {et~al.},\ }\bibfield  {title} {\bibinfo
  {title} {Evidence for topological magnon–phonon hybridization in a {2D}
  antiferromagnet down to the monolayer limit},\ }\href
  {https://doi.org/10.1021/acs.nanolett.3c00351} {\bibfield  {journal}
  {\bibinfo  {journal} {Nano Letters}\ }\textbf {\bibinfo {volume} {23}},\
  \bibinfo {pages} {2023} (\bibinfo {year} {2023})}\BibitemShut {NoStop}%
\bibitem [{\citenamefont {Mai}\ \emph {et~al.}(2021)\citenamefont {Mai},
  \citenamefont {Garrity}, \citenamefont {McCreary}, \citenamefont {Argo},
  \citenamefont {Simpson}, \citenamefont {Doan-Nguyen}, \citenamefont
  {Aguilar},\ and\ \citenamefont {Walker}}]{mai2021magnon}%
  \BibitemOpen
  \bibfield  {author} {\bibinfo {author} {\bibfnamefont {T.~T.}\ \bibnamefont
  {Mai}}, \bibinfo {author} {\bibfnamefont {K.~F.}\ \bibnamefont {Garrity}},
  \bibinfo {author} {\bibfnamefont {A.}~\bibnamefont {McCreary}}, \bibinfo
  {author} {\bibfnamefont {J.}~\bibnamefont {Argo}}, \bibinfo {author}
  {\bibfnamefont {J.~R.}\ \bibnamefont {Simpson}}, \bibinfo {author}
  {\bibfnamefont {V.}~\bibnamefont {Doan-Nguyen}}, \bibinfo {author}
  {\bibfnamefont {R.~V.}\ \bibnamefont {Aguilar}},\ and\ \bibinfo {author}
  {\bibfnamefont {A.~R.~H.}\ \bibnamefont {Walker}},\ }\bibfield  {title}
  {\bibinfo {title} {Magnon-phonon hybridization in {2D} antiferromagnet
  {MnPSe$_3$}},\ }\href {https://doi.org/10.1126/sciadv.abj3106} {\bibfield
  {journal} {\bibinfo  {journal} {Science Advances}\ }\textbf {\bibinfo
  {volume} {7}},\ \bibinfo {pages} {eabj3106} (\bibinfo {year}
  {2021})}\BibitemShut {NoStop}%
\bibitem [{\citenamefont {Harvey}(1933)}]{harvey1933diffuse}%
  \BibitemOpen
  \bibfield  {author} {\bibinfo {author} {\bibfnamefont {G.}~\bibnamefont
  {Harvey}},\ }\bibfield  {title} {\bibinfo {title} {{Diffuse scattering of
  X-rays from Sylvine. IV. Scattering at high temperatures}},\ }\href
  {https://doi.org/10.1103/PhysRev.44.133} {\bibfield  {journal} {\bibinfo
  {journal} {Physical Review}\ }\textbf {\bibinfo {volume} {44}},\ \bibinfo
  {pages} {133} (\bibinfo {year} {1933})}\BibitemShut {NoStop}%
\end{thebibliography}%

\end{document}